\documentclass[final,3p,times,authoryear]{elsarticle}

\usepackage{graphicx}
\usepackage{amssymb}
\usepackage{amsmath}
\usepackage{float}

\usepackage{mathtools}
\usepackage[colorlinks]{hyperref}
\usepackage{cleveref}

\usepackage{xcolor}

\newcommand{\overbar}[1]{\mkern 1.5mu\overline{\mkern-1.5mu#1\mkern-1.5mu}\mkern 1.5mu}\newcommand{\underbarnew}[1]{\mkern 1.5mu\underline{\mkern-1.5mu#1\mkern-1.5mu}\mkern 1.5mu}

\usepackage{paralist}
\usepackage{enumitem}
\usepackage{lineno}

\journal{Energy Economics}

\begin{document}

\begin{frontmatter}


\title{On Wholesale Electricity Prices and Market Values in a Carbon-Neutral Energy System}

\author{Diana Böttger, Philipp Härtel}

\address{Fraunhofer Institute for Energy Economics and Energy System Technology IEE, Königstor 59, 34119 Kassel, Germany, diana.boettger@iee.fraunhofer.de}

\begin{abstract}

Climate and energy policy targets of the European Commission aim to make Europe the first climate-neutral continent by 2050.
For low-carbon and net-neutral energy systems primarily based on variable renewable power generation, issues related to the market integration, cannibalisation of revenues, and cost recovery of wind and solar photovoltaics have become major concerns.
The traditional discussion of the merit-order effect expects wholesale power prices in a system with 100\,\% renewable energy sources to alternate between very high and very low values.
Unlike previous work, we present a structured and technology-specific analysis of the cross-sectoral demand bidding effect for the price formation in low-carbon power markets.
Starting from a stylised market arrangement and by successively augmenting it with all relevant technologies, we construct and quantify the cross-sectoral demand bidding effects in future European power markets with the cross-sectoral market modelling framework SCOPE SD. 
As the main contribution, we explain and substantiate the market clearing effects of new market participants in detail. 
Hereby, we put a special focus on hybrid heat supply systems consisting of combined heat and power power plant, fuel boiler, thermal storage and electrical back up and derive the opportunity costs of these systems.
Furthermore, we show the effects of cross-border integration for a large-scale European net-neutral energy scenario.
Finally, the detailed information on market clearing effects allows us to evaluate the resulting revenues of all major technology categories on future electricity markets. 
\end{abstract}

\begin{keyword}
Low-carbon energy system \sep electricity price  \sep energy system modelling \sep merit-order effect \sep multivalent CHP
\end{keyword}

\end{frontmatter}

\section{Motivation}
\label{S:Motivation}

The European Commission aims to make Europe the first climate-neutral continent by 2050~\citep{EuropeanCommission.11December2019}, requiring the large-scale decarbonisation of power, building, industry, and transport sector.
Adopting the climate neutrality paradigm and bringing net greenhouse gas (GHG) emissions to zero has far-reaching consequences for the economy and renders prospects for business cases based on fossil fuels increasingly difficult.

Wind and solar photovoltaics (PV) are assumed to become the primary source of energy, accompanied by hydropower and, to some extent, geothermal energy and biomass.

\subsection{Merit-order effect in short to medium-term systems}

It is well known that increased amounts of variable renewable energy sources (RES) in the power system depress market clearing prices through the so-called `merit-order effect', e.g. see~\citep{Sensfu.2008,Cludius.2014,Ketterer.2014,Ackermann.2015,Bird.2016,Djrup.2018,Brown.2020,Halttunen.2020}.

The number of hours with zero market clearing prices due to curtailment is often expected to increase.
Due to a combination of factors such as congestion, start-up and shut-down costs of thermal plants, and fixed feed-in tariffs for RES, several markets are already today experiencing some periods with zero or negative prices~\citep{Ketterer.2014,Ackermann.2015,Bird.2016,Djrup.2018,Halttunen.2020}.
Many occurrences of zero and negative prices should be possible to be avoided by better grid planning, better flexibility products in the market, and fully exposing wind and solar PV production to the markets.
Nevertheless, curtailment is inevitable when the RES output exceeds the demand, resulting in market clearing prices equal to zero if the system does not have sufficient demand flexibility or storage capacity available.
On the other hand, extremely high prices can be expected more often in the periods of low solar PV and wind output if the base and mid-merit plants are pushed out of the market due to the merit-order effect~\citep{Winzer.2019}.

\subsection{Merit-order effect in low-carbon systems}
Regarding low-carbon and net-neutral energy systems primarily based on power generation from variable renewable energy generation, market integration, cannibalisation of revenues, and cost recovery of wind and solar PV have become major concerns~\citep{Mills.2015,Brown.2020}.

A hypothetical scenario with 100\,\% variable renewable supply is described in~\citep{Joskow.2019}, which leads to a situation where the prices alternate between zero and a price cap depending on the net load.

Moreover, the analysis in~\citep{Panos.2019} finds that there is a clear shift towards more zero and high price periods in a low-carbon than a baseline scenario for Europe in 2030.
Similar results can be found e.g. in~\citep{Ketterer.2014,Ackermann.2015,Bird.2016,Djrup.2018}.

Compared to the possible import or domestic generation of synthetic hydrocarbon fuels, which can be used for different purposes, direct electrification of final energy demands seems to be the most cost-efficient decarbonisation measure in many cases, if feasible.

Consequently, cross-sectoral interactions become more important, and the overall electricity demand is expected to rise.
Flexible power demand from the heat and transport sector can enhance the integration of variable RES, particularly by bi- and multivalent technology applications.
Beyond that, the discussion focus shifts away from the much-discussed integration of renewable energies towards identifying new ways of using variable renewable electricity production in different end-use sectors.
The potential flexibility of cross-sectoral integration at the centralised and decentralised level of the system is generally not sufficiently taken into account in the above mentioned literature.

\subsection{Cross-sectoral demand bidding}
The work in~\citep{Hartel.2021} demonstrates that the explicit modelling and quantification of cross-sectoral demand bidding are crucial factors in the electricity price formation of low-carbon electricity markets.
It notes that there is only a limited role of zero or even negative price situations due to excess renewable production since hybrid cross-sectoral consumers with valid opportunity costs will use it to supply final heat, industry, and transport demands.
These mainly involve heating sector technologies, i.e. hybrid boiler and multivalent combined heat and power (CHP) systems, as they present the lowest marginal value when compared to the considered hybrid transport sector applications.

The study in \citep{Bernath.2021} analyses a carbon-neutral German energy system within Europe with different cross-sectoral flexibility options. They find that providing cross-sectoral flexibility reduces curtailment of RES and increases the market values of onshore wind and solar PV. Their analysis encompasses flexibility from electric vehicles, heat pumps with thermal storage as well as multivalent CHP systems incl. thermal storage, fuel boiler and electric boiler. Flexible electrolysis is not considered. Furthermore, they derive power prices as shadow prices from the dispatch and investment decision optimization problem. This means that their power prices deviate from a purely dispatch optimization in hours where the investment decision variables are price setting. 

\par
Unpublished work in~\citep{Ruhnau.2020} reasons why wind and solar market values stabilise in long-term markets due to flexible demand from hydrogen electrolysis. 
By only focusing on electrolyser technology, the analysis does not include (hybrid) technology combinations that might enter electricity markets earlier than electrolyser units.
Note that the role of hydrogen electrolysis in the price formation of future electricity markets is also subject to whether low-carbon hydrogen or any derived hydrocarbons are produced domestically or imported from outside the bidding zone configuration of the considered electricity market.

\subsection{Research question}

Since there are concerns that wholesale electricity market prices in a power system with 100\,\% RES are either very high or very low due to the merit-order effect~\citep{Sensfu.2008, Cludius.2014, Praktiknjo.2016, PaulJohnson.2019}, we are going to analyse the effect of cross-sectoral demand bidding for the price formation in future electricity markets.

Hybrid consumer applications with true opportunity costs, e.g. for backup fuel (of fossil or renewable origin), could make demand bids in times of purely renewable electricity supply, i.e. marginal production cost close to zero, which introduces new plateaus in the market price duration curve.
In contrast to previous work \citep{Hartel.2018c,Hartel.2021}, which analysed a -\,87.5\,\% GHG emission reduction scenario, we explore a carbon net-neutral target scenario for Europe.
Furthermore, by using a somewhat stylised scenario setup for a single bidding zone, using Germany as an example, we first want to dissect and explain the different steps in the low-carbon electricity market's price duration curve in detail.

The SCOPE SD modelling and optimisation framework, which is developed and maintained at Fraunhofer~IEE, captures cross-sectoral interactions and a variety of technology combinations when analysing future electricity market clearing and facilitates the underlying case study.

\subsection{Paper structure}

The remainder is organised as follows:
\Cref{S:Methodology} explains the methodology including the modelling and optimisation approach to investigate wholesale electricity prices in net-neutral energy systems.
\Cref{S:Casestudy} sets out the structure and assumptions of the case study conducted for a carbon-neutral energy system in Europe.
\Cref{S:Results} presents and section \ref{S:Discussion} discusses the case study simulation results.
\Cref{S:SummaryConclusion} closes with a summary and draws relevant conclusions.

\section{Methodology}
\label{S:Methodology}

\subsection{Model overview}

We use the cross-sectoral capacity expansion planning framework ``SCOPE Scenario Development'' (SCOPE SD) \citep{SCOPEFlyer}
to develop a long-term carbon-neutral energy system scenario in Europe for the scenario year 2050.
The underlying large-scale optimisation model captures the power, building, industry, and transport sector for each country modelled as one bidding zone.
By meeting climate targets as well as the end-use demands of each sector in every hour of the considered scenario year, the cross-sectoral capacity expansion planning model is able to determine the cost-optimal installed capacities including the economic dispatch of single technologies.
\par
Importantly, the modelling approach allows for the extraction of approximate wholesale electricity market prices, i.e. shadow prices, from the dual variables of the market clearing constraint of the large-scale LP problem instances.
Since the analysed scenario setting only features a few remaining thermal power plants and integer variables, unit commitment decisions become less critical for the overall objective and the LP approach seems to be justified.
As a side benefit, we can also bypass the problem of non-existent dual variables in mixed-integer programming problems~\citep{ONeill.2005,Huppmann.2018,Hartel.2021}.
\par
The analysed scenario setting assumes the existence of hybrid technologies which can offer their flexibility to the electricity market. To be able to see the market clearing effects, it is necessary to model the bi- and multivalent technologies explicitly.
For example, these include complex systems consisting of combined heat and power plant, thermal backup boiler, electric backup boiler (i.e. direct resistive heating or heat pump), and thermal storage units. 

\begin{figure*}
    \centering
    \includegraphics[width=\linewidth]{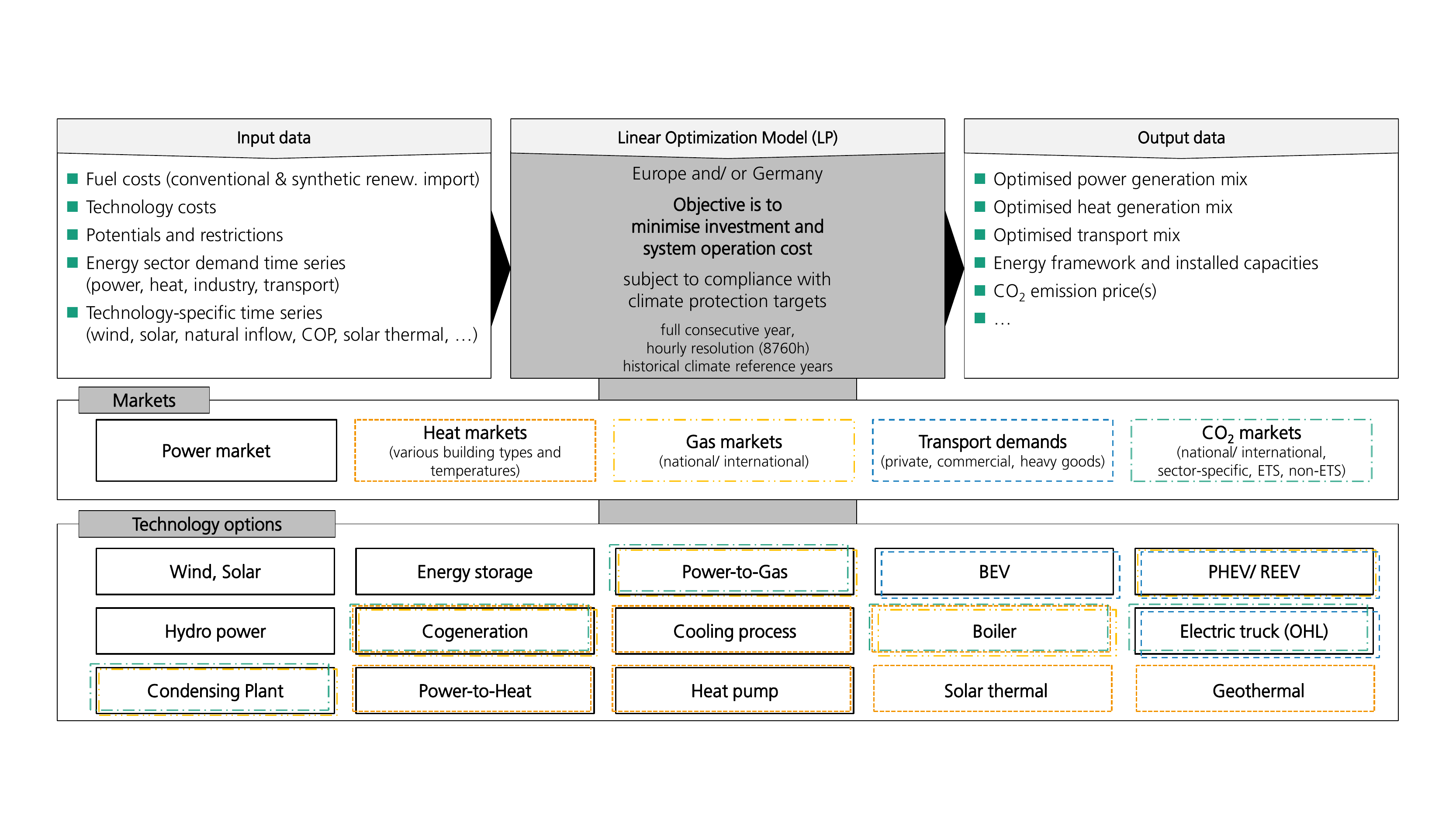}
    \caption{Overview of the modelling and optimisation framework SCOPE Scenario Development (SD), own illustration.}
    \label{fig:SCOPESDoverview}
\end{figure*}

Fig.~\ref{fig:SCOPESDoverview} shows a schematic illustration of the SCOPE~SD structure as well as its in- and outputs.

\subsection{Modelling of technologies}

To be able to gain the desired insights, it is essential to explain the modelling approach and its sub-modules.
Therefore, a detailed description of the employed energy system model follows below, including the individual technologies considered for the power, heat, and transport sector.

\par
Power generating technologies include variable RESs, i.e. wind and solar PV, as well as hydro-thermal power plants. Power generation from biomass is not considered in the underlying case study setting.
Power-to-gas plants are power consumers and batteries are both, generators and consumers.
The end-use demand in the heating sector is supplied by various technologies, including CHP units, fuel boilers, electric backup units (direct resistive heating or heat pump), and thermal storage systems.
Cooling systems for air conditioning are modeled with a thermal storage.
Electric vehicles may contribute flexibility through controlled charging and use of their battery storage as well as multiple-fuel flexibility for hybrid electric vehicles.

\par
Conceptionally, the SCOPE~SD model can optimise investment and dispatch decisions for all sectors in a co-optimisation approach.
Note that the analysis of the case study scenario variants primarily uses the economic dispatch configuration of the modelling and optimisation framework.
However, to maintain a consistent net-neutral energy framework for the European case study, these variants are based on a preceding run of the full cross-sectoral capacity expansion planning configuration, see \Cref{S:Casestudy}.

\par
With a single-period planning horizon, the model optimises one future target year, here 2050, representing the full chronological year in hourly resolution.
Individual time steps are denoted as $t\in T$.
Note that all following decision variables are non-negative, i.e. $\geq 0$.

\subsubsection{Variable renewable power generation}

Variable renewable power generation $r\in R$ such as on- and offshore wind power, as well as solar PV plants are able to produce electricity following a predefined hourly availability time series $AV_{r,t} \in [0,1]$.
The modelling approach allows for market-based curtailment $x^\text{CU}_{r,t}$, which is compensated with variable cost savings $C^\text{CU}_{r}$.
With the predefined installed capacity $\overbar{X}^\text{GEN}_{r}$, the renewable generation $x^\text{GEN}_{r,t}$ is limited by \Cref{eq:genLimitsRen}, and contributes to the objective function with $\sum_{r\in R} C^\text{VP}_{r} x^\text{GEN}_{r,t} + C^\text{CU}_{r} x^\text{CU}_{r,t}$ with variable production costs $C^\text{VP}_{r}$.

\begin{equation}
0 \leq x^\text{GEN}_{r,t} + x^\text{CU}_{r,t}\leq AV_{r,t}  \overbar{X}^\text{GEN}_{r}  
\qquad \forall r \in R, \quad \forall t
\label{eq:genLimitsRen}
\end{equation}

\subsubsection{Hydropower systems}

Hydropower system units $u\in U$ are modelled with turbines, pumped-turbines, pumps, as well as their storage reservoirs.
By using the equivalent reservoir approach developed in \citep{Hartel.2017}, the hydropower reservoir formulations distinguish a main reservoir and a potentially additional synthetic pumped-hydro reservoir to approximate the operational characteristics of complex multireservoir systems.
Data for (equivalent) hydropower systems describes equivalent natural energy inflow profiles into the main reservoir $X^\text{N}_{u,t}$ as well as corresponding equivalent natural energy inflow volumes into the pumped-hydro reservoir $X^\text{N,P}_{u,t}$.
Furthermore, the power plants are characterised by maximum turbine generation limits $\overbar{X}^\text{GEN,T}_{u,t}$, maximum pumped-turbine generation limits $\overbar{X}^\text{GEN,PT}_{u,t}$, maximum pump consumption limits $\overbar{X}^\text{CON}_{u,t}$, as well as the equivalent pump efficiency factor $\eta^\text{P}_u$.
Decision variables are turbine generation $x^\text{GEN,T}_{u,t}$, pumped-turbine generation $x^\text{GEN,PT}_{u,t}$, pump consumption $x^\text{CON}_{u,t}$, conventional reservoir storage level $x^\text{S}_{u,t}$ and pumped reservoir storage level $x^\text{S,P}_{u,t}$. Additionally, the variables for conventional reservoir spillage $x^\text{SP}_{u,t}$ and pumped reservoir spillage $x^\text{SP,P}_{u,t}$ ensure the feasibility of the problem.
\par
Hydropower generation is limited by \Cref{eq:genLimitsHydro} and potential pump consumption by \Cref{eq:conLimitsHydro}.

\begin{align}
    &\begin{multlined}
    x^\text{GEN}_{u,t} = x^\text{GEN,T}_{u,t} + x^\text{GEN,PT}_{u,t}\quad \forall u, t \quad
    \text{ with } x^\text{GEN,T}_{u,t} \leq \overbar{X}^\text{GEN,T}_{u,t}\,,
    \quad x^\text{GEN,PT}_{u,t} \leq \overbar{X}^\text{GEN,PT}_{u,t}\,
    \label{eq:genLimitsHydro}
    \end{multlined}\\
    & x^\text{CON}_{u,t} \leq \overbar{X}^\text{CON}_{u,t}\,\quad \forall u, t
    \label{eq:conLimitsHydro}
\end{align}

The storage continuity conditions for the equivalent main and pump-hydro reservoirs are given in \Cref{eq:hydroStorageBalance,eq:hydroPumpedStorageBalance}, respectively, see also \citep{Hartel.2017}.
The considered hydropower systems have no explicit cost contributions to the objective function of the problem, see \citep{Hartel.2021a}.

\begin{align}
    &\begin{multlined}
    x^\text{S}_{u,t+1} = x^\text{S}_{u,t} + X^\text{N}_{u,t} - x^\text{GEN,T}_{u,t} - x^\text{GEN,PT}_{u,t} + x^\text{CON}_{u,t} \, \eta^\text{P}_u - x^\text{SP}_{u,t} - x^\text{SP,P}_{u,t} \quad \forall u, t\label{eq:hydroStorageBalance}
    \end{multlined}\\
    &\begin{multlined}
    x^\text{S,P}_{u,t+1} = x^\text{S,P}_{u,t} + X^\text{N,P}_{u,t} - x^\text{GEN,PT}_{u,t} + x^\text{CON}_{u,t} \, \eta^\text{P}_u - x^\text{SP,P}_{u,t} \quad \forall u, t\label{eq:hydroPumpedStorageBalance}
    \end{multlined}
\end{align}

\subsubsection{Thermal power plants and gas turbines}
\label{S:GTmodeling}

In the analysed scenario, only-open cycle gas turbines (OCGTs) are relevant as flexible (backup) power plants and neither combined-cycle gas turbines (CCGT) nor coal power plants serve as firm generation capacity.
In a few countries, nuclear power plants have not reached the end of their estimated lifetime and, therefore, remain in the thermal generation stacks.
Moreover, combined heat and power (CHP) systems include CCGT or OCGT plants which may partly assume the role of flexible generation, which will be further explained in \Cref{S:CHPmodeling}.
Note that all variables and parameters for modelling both CHP and non-CHP units are explained in this section.
\par
All conventional generation units $g\in G$ are characterised by their installed power generation capacity $\overbar{X}^\text{GEN}_{g}$.
Due to the representation of technology clusters in each market area and the resulting linear programming approach for the problem, the minimum power generation limit $\underbarnew{X}^\text{GEN}_{g,t}$ is set to 0.
The hourly availability $AV_{g,t}$ in\,\% of installed capacity describes the technical hourly availability which follows a seasonal pattern, limiting the power output $x^\text{GEN}_{g,t}$ in \Cref{eq:genLimits}.
To restrict arbitrary load changing of single power plant clusters in the continuous model without unit commitment restrictions, positive and negative load change variables $x^{\text{LC}+/-}_{g,t}$ keep track of load cycling (LC) in \Cref{eq:genLoadChange}.

\begin{align}
    0=\underbarnew{X}^\text{GEN}_{g,t} \leq x^\text{GEN}_{g,t} \leq AV_{g,t} \,  \overbar{X}^\text{GEN}_{g} \quad \forall g \in G, \forall t\label{eq:genLimits}\\
    x^\text{GEN}_{g,t+1} = x^\text{GEN}_{g,t} + x^{\text{LC}+}_{g,t} - x^{\text{LC}-}_{g,t} \quad \forall g \in G \setminus G^\text{CHP}, \forall t \label{eq:genLoadChange}
\end{align}

Both positive and negative load changes are penalised in the objective function by the term
\begin{equation}
\sum_{t\in T} \Bigg ( \sum_{g \in G} \bigg( C_g^\text{VP}\,x^\text{GEN}_{g,t} + C_g^\text{LC} \left( x^{\text{LC}+}_{g,t}+x^{\text{LC}-}_{g,t} \right) \bigg)\,,
\end{equation}
where $C_g^\text{VP}$ are variable production costs and $C_g^\text{LC}$ represent ramping costs.

\subsubsection{Power-to-Gas plants}

We model power-to-gas plants $l\in L$ as flexible power consumers with a power-to-fuel conversion factor $PF_{l}$ measured in MWh\textsubscript{th}/MWh\textsubscript{el}.
The variable $y_{l,t}^{\text{GEN}}$ describes the fuel generation, see \Cref{eq:powerToFuelLimits}, which is compensated by the methane price $P^{CH_4}$ excluding national transport costs in the objective function.

\begin{equation}
 y^\text{GEN}_{l,t} = PF_{l} x^\text{CON}_{l,t}  \qquad \forall l \in L, \quad \forall t\label{eq:powerToFuelLimits}
\end{equation}

The power consumption $x^\text{CON}_{l,t}$ is limited by the predefined installed capacity.
Similar to thermal power plants, the positive and negative load change $x^{\text{LC}+/-}_{l,t}$, recall \Cref{eq:conLoadChange}, is penalised with load change cost $C_{l}^\text{LC}$ in the objective function by the following term
\begin{equation}
\sum_{t\in T} \Bigg ( \sum_{l \in L} \bigg( -P^{CH_4} y^\text{GEN}_{l,t} +  C_l^\text{LC} \left( x^{\text{LC}+}_{l,t}+x^{\text{LC},-}_{l,t} \right)\bigg)\bigg)\,.
\end{equation}
Additional costs for the power consumption are not modelled.

\begin{equation}
    x^\text{CON}_{l,t+1} =  x^\text{CON}_{l,t} + x^{\text{LC}+}_{l,t} - x^{\text{LC}-}_{l,t} 
    \qquad \forall l \in L, \quad \forall t\label{eq:conLoadChange}
\end{equation}

\subsubsection{Stationary electricity storage systems}

Stationary (battery) storage system units $a\in A$ are modelled with
storage input efficiency $\eta^\text{IN}_{a}$, storage output efficiency $\eta^\text{OUT}_{a}$, linear electricity storage loss factor $\lambda_{a}^\text{S}$ and linear electricity storage self-discharge factor $\Lambda_{a}^\text{S}$. The model optimizes electricity storage input $x^\text{IN}_{a,t}$, output $x^\text{OUT}_{a,t}$ and the electricity storage level $x^\text{S}_{a,t}$ according to the storage continuity of electricity storage in \Cref{eq:electricityStorageContinuity}.

\begin{equation}
    x^{\text{S}}_{a,t+1}= x^{\text{S}}_{a,t}  (1-\lambda^{\text{S}}_{a}) - \Lambda^{\text{S}}_{a} + x_{a,t}^{\text{IN}} \eta^{\text{IN}}_{a}
        - \tfrac{x_{a,t}^{\text{OUT}}}{\eta^{\text{OUT}}_{a}} \quad \forall a, t\label{eq:electricityStorageContinuity}
\end{equation}

Again, the installed capacity is predefined and load change costs are considered in the objective function with $C_a^\text{LC} \left( x^{\text{LC}+}_{a,t}+x^{\text{LC}-}_{a,t} \right)$ neglecting further power consumption costs.

\subsubsection{Heat supply systems}

The heating sector is considered in form of different heat markets $m\in M^\text{HEAT}$, e.g. for the supply of households, the trade and service sector, and the industry.
We distinguish between onsite heat supply and district heating networks.
Note that the installed installed capacities of all heat supply technologies are fixed and only their dispatch is optimised.
There are always hybrid systems installed which can have the following form, see also \citep{Hartel.2021}:
\begin{itemize}
    \item \textbf{Multivalent CHP systems}, consisting of a CHP unit (extraction condensing CCGT or OCGT with waste heat boiler), a condensing boiler unit, an electric backup unit (i.e. direct resistive heating or heat pump), a thermal storage unit, and an optional solar thermal unit;
    \item \textbf{Hybrid boiler systems}, including a condensing boiler unit, an electric backup boiler unit, and an optional solar thermal unit;
    \item \textbf{Hybrid heat pump systems}, with a heat pump as the main unit, a potential electric or fuel boiler as backup unit, and a thermal storage unit.
\end{itemize}

\paragraph{Multivalent CHP systems}
\label{S:CHPmodeling}

Additionally to the described modelling of thermal power plants in subsection \ref{S:GTmodeling} the variable $q^\text{CHP}_{g,t}$ describes the heat generation for all CHP system units $g\in G^\text{CHP} \subseteq G$. CHP units are further characterised by their power-to-heat ratio (backpressure limit) factor $PH_{g}$ and power loss factor $PL_{g}$.

The load change equation in \Cref{eq:genLoadChange} is modified for CHP units to incorporate the heat generation in \Cref{eq:chpLoadChange}.

\begin{align}
\begin{split}
    x^{\text{LC}-}_{g,t} - x^{\text{LC}+}_{g,t} = &x^\text{GEN}_{g,t}-x^\text{GEN}_{g,t+1} +PL_g \left( q^\text{CHP}_{g,t}- q^\text{CHP}_{g,t+1}\right)  \quad \forall g \in G^\text{CHP}, \quad \forall t \in T\,. 
    \label{eq:chpLoadChange}
\end{split}
\end{align}

Modelling of the CHP plant further takes into account the characteristic linearised CHP diagram in \Cref{eq:chpLimits1,eq:chpLimits2}.
Each CHP plant system comprises also a condensing boiler which is characterised by its  efficiency $\eta^\text{CB}_{g}$. The condensing boiler fuel consumption $y^\text{CB}_{g,t}$ is limited by the design factor of the CHP system $\Pi^\text{BU}_g$ in \Cref{eq:chpBoilerDimensioning}. Power consumption of backup electrical heaters in CHP systems $x^\text{CON}_{g,t}$ is restricted according to \Cref{eq:chpHeaterDimensioning} taking into account the design factor of the CHP system $\Pi^\text{CON}_g$.

\begin{align}
    x^\text{GEN}_{g,t} + PL_g  q^\text{CHP}_{g,t} \leq AV_{g,t} 
    \overbar{X}^\text{GEN}_{g}
    \quad \forall g \in G^\text{CHP}, \forall t\label{eq:chpLimits1}\\
    0 \leq PH_g  q^\text{CHP}_{g,t} \leq x^\text{GEN}_{g,t} \quad \forall g \in G^\text{CHP}, \forall t\label{eq:chpLimits2}\\
    0 \leq \eta^{\text{CB}}_{g} \, y^\text{CB}_{g,t} \leq \Pi^\text{BU}_g \overbar{X}^\text{GEN}_{g} \quad \forall g \in G^\text{CHP}, \forall t\label{eq:chpBoilerDimensioning}\\
    0 \leq x^\text{CON}_{g,t} \leq \Pi^\text{CON}_g \, \overbar{X}^\text{GEN}_{g} \quad \forall g \in G^\text{CHP}\label{eq:chpHeaterDimensioning}
\end{align}

Thermal storage in the heat supply system is characterised by its input efficiency $\eta^\text{IN}_{g}$ and output efficiency $\eta^\text{OUT}_{g}$.
Variables of the thermal storage are storage level $q^\text{S}_{g,t}$, storage input $q^\text{IN}_{g,t}$ and storage output $q^\text{OUT}_{g,t}$.
The thermal storage continuity is given in \Cref{eq:thermalStorageContinuity} analogously to stationary storage systems.

\begin{equation}
\begin{multlined}
    q^{\text{S}}_{g,t+1}= q^{\text{S}}_{g,t} \, (1-\lambda^{\text{S}}_{g}) - \Lambda^{\text{S}}_{g} + q_{g,t}^{\text{IN}} \,\eta^{\text{IN}}_{g}
        - \tfrac{q_{g,t}^{\text{OUT}}}{\eta^{\text{OUT}}_{g}} \quad \forall g \in G^\text{CHP}, \quad \forall t \in T
    \label{eq:thermalStorageContinuity}
\end{multlined}
\end{equation}

The heat demand balance of CHP systems with heat demand $D_{m,t}$ is described in \Cref{eq:chpDemandCoverage} with a linear heating network transmission loss factor $\chi_{g}$ and a solar thermal contribution factor $ST_{g,t} \in [0,1]$.
Solar thermal plant units are modelled with a weather-dependent hourly solar thermal generation profile. The efficiency of the electric backup is given by $\eta^\text{CON}_g$.

\begin{equation}
\begin{multlined}
    \tfrac{ST_{g,t} \, D_{m,t}}{1-\chi_{g}} = q^{\text{CHP}}_{g,t} + \eta^{\text{CB}}_{g} \, y^{\text{CB}}_{g,t} + \eta^\text{CON}_g \, x^{\text{CON}}_{g,t} + q^{\text{OUT}}_{g,t} - q^{\text{IN}}_{g,t} \quad
    \forall g\in G^\text{CHP}_m, \forall m\in M^\text{HEAT}, \forall t\in T \label{eq:chpDemandCoverage}
\end{multlined}
\end{equation}

The costs of CHP plants in the objective function encompass variable costs, load change costs and costs for the heat generation as well as boiler fuel costs in \Cref{eq:chpCosts}.

\begin{equation}
\begin{multlined}
    \sum_{g\in G^\text{CHP}}  C_g^\text{VP} x^\text{GEN}_{g,t} + C_g^\text{LC} \left( x^{\text{LC}+}_{g,t}+x^{\text{LC},-}_{g,t} \right)
    +C_g^\text{CHP} q^\text{CHP}_{g,t} + C_g^\text{CB} y^\text{CB}_{g,t}
    \label{eq:chpCosts}
\end{multlined}
\end{equation}

Here is $C_g^\text{VP} = \frac{P^{CH_4}}{\eta^\text{GEN}_g} $ and $C_g^\text{CHP} = P^{CH_4} \cdot PL_{g} $. 

There are no variable costs for the electrical backup unit aside from the necessary power generation costs on the market.

\paragraph{Hybrid boiler systems}

Hybrid boiler systems $b\in B$ encompass a condensing boiler as well as an  electrical backup. Condensing boilers are again characterised by their efficiency $\eta^\text{CB}_{b}$ and variable production cost $C_{b}^\text{CB}$. The condensing boiler's fuel consumption $y^\text{CB}_{b,t}$ enters the objective function with $\sum_{b\in B}  C_b^\text{CB} y^\text{CB}_{b,t}$. \Cref{eq:hbDemandCoverage} describes the heat supply balance. The linear heating network transmission loss factor $\chi_{b}$ equals zero for on-site generation and is greater than zero for district heating grids.

\begin{equation}
\begin{multlined}
    \tfrac{ST_{b,t}  \, D_{m,t}}{1-\chi_{b}} = \eta^{\text{CB}}_{b} \, y^{\text{CB}}_{b,t} + \eta^\text{CON}_b \, x^{\text{CON}}_{b,t} \quad \forall b\in B_m, \quad \forall m\in M^\text{HEAT},\quad \forall t\in T\,.
    \label{eq:hbDemandCoverage}
\end{multlined}
\end{equation}

\paragraph{Hybrid heat pump systems}

Heat pump systems $h\in H$ consist of a heat pump as the main unit, a potential electric $\left (x^{\text{CON,BU}}_{h,t}\right )$ or fuel boiler $\left (y^{\text{CON}}_{h,t}\right )$ backup unit, as well as a thermal storage unit $\left (q^{\text{IN}}_{h,t}\text{ and } q^{\text{OUT}}_{h,t}\right )$.
The modelling approach corresponds to the lumped modelling approach presented in \citep{Hartel.2020b}, i.e. individual end-use heating demands are considered in a combined manner.
For instance, space heating and hot water heating in the building sector are considered with a `lumped' demand profile.
The heat pump unit is mainly characterised by the electricity consumption $x^{\text{CON,HP}}_{h,t}$ and its coefficient of performance (COP) $\eta^\text{CON}_{h,t}$, which indicates the thermodynamic efficiency as the ratio of the useful heating or cooling energy provided and the input energy consumed by the heat pump.
Note that the technology-specific profile $\eta^\text{CON}_{h,t}$ varies with time since it depends on both the temperature level of the external heat source (air, ground, water) and the required end-use temperature of the heat sink, see \citep{Hartel.2020b} for a detailed description of the time-series data generation.
\par
The total power consumption of this heat supply system equals
\begin{equation}
x^\text{CON}_{h,t} = x^{\text{CON,HP}}_{h,t}+x^{\text{CON,BU}}_{h,t} \quad \forall h,t\,.
\end{equation}

The thermal storage is restricted according to 
\begin{equation}
0\leq q^{\text{IN}}_{h,t} \leq \eta^\text{CON}_{h,t}\,x^{\text{CON,HP}}_{h,t} \quad \forall h,t\,.
\end{equation}

\Cref{eq:heatPumpDemandCoverage} represents the heat demand coverage of the hybrid heat pump system.

\begin{equation}
\begin{multlined}
    ST_{h,t} \, D_{m,t} = \eta^\text{CON}_{h,t}\,x^\text{CON,HP}_{h,t} 
 + \eta^{\text{CON}}_{h}\,x^{\text{CON,BU}}_{h,t} + \eta^{\text{CB}}_{h} \, y^\text{CON}_{h,t} +q^{\text{OUT}}_{h,t}- q^{\text{IN}}_{h,t} \quad \forall h\in H_m, \quad \forall m\in M^\text{HEAT}, \quad \forall t\in T     \label{eq:heatPumpDemandCoverage}
\end{multlined}
\end{equation}

\subsubsection{Air conditioning}

Cooling system units $o\in O$ are characterised by their electric efficiency $\eta^\text{CON}_{o}$ and have to fulfill a predefined cooling demand profile $D_{m,t}$, see \Cref{eq:coolingDemandCoverage}.
The variable $x^\text{CON}_{o,t}$ represents the power consumption for air conditioning.
Potential operational flexibility of cooling systems is incorporated by modeling a thermal storage unit in the cooling demand balance defined by \Cref{eq:thermalStorageContinuityAirCond}.

\begin{gather}
    \begin{multlined}
    D_{m,t} = \eta^\text{CON}_{o,t}  x^\text{CON}_{o,t} - q^\text{IN}_{o,t} + q^\text{OUT}_{o,t}  \quad \forall o\in O_m, \quad 
    \forall m\in M^\text{COOL},\quad \forall t\in T
    \label{eq:coolingDemandCoverage}
    \end{multlined}\\
    \begin{multlined}
    q^{\text{S}}_{o,t+1}= q^{\text{S}}_{o,t} \, (1-\lambda^{\text{S}}_{o}) - \Lambda^{\text{S}}_{o} + q_{o,t}^{\text{IN}} \,\eta^{\text{IN}}_{o}
        - \tfrac{q_{o,t}^{\text{OUT}}}{\eta^{\text{OUT}}_{o}} \quad \forall o\in  O,\quad \forall t\in T \label{eq:thermalStorageContinuityAirCond}
    \end{multlined}
\end{gather}

\subsubsection{Transport sector}

The modeling of the transport sector\footnote{See also \citep{Boettger.2018} for more details concerning the modeling of the transport sector.} encompasses different road transport markets $m\in M^\text{ROAD}$ and electric vehicle units $v\in V$ like battery electric vehicles (BEV) or plug-in hybrid vehicles (PHEV). Hybrid vehicles can choose between driving with internal combustion engine or electric engine on an hourly time scale. The variable $z_{v,t}^{\text{ICE}}$ describes the distance driven by internal combustion engine and $z_{v,t}^{\text{EL}}$ the distance driven by electric drive  in the transport demand balance in \Cref{eq:vehicleDemandCoverage} where $\phi_{v}$ is the market share of vehicle type $v$ and $D_{m,t}$ is the driving demand.

\begin{equation}
    \phi_{v} \, D_{m,t} = z^{\text{ICE}}_{v,t} + z^{\text{EL}}_{v,t} \quad \forall v\in V_m, \forall m\in M^\text{ROAD}, \forall t\label{eq:vehicleDemandCoverage}
\end{equation}

We differentiate between controlled and uncontrolled charging where $FS_{v} \in [0, 1]$ is the flexible charging share. The electric and conventional driving limits in \Cref{eq:vehicleElectricDrivingLimit,eq:vehicleIceDrivingLimit} have to consider the maximum distance driven by electric drive $\overline{Z}^{\text{EL}}_{v,t}$.

\begin{gather}
    FS_{v} \, \overline{Z}^{\text{EL}}_{v,t} \, \phi_{v} \leq z^{\text{EL}}_{v,t} \leq \overline{Z}^{\text{EL}}_{v,t} \, \phi_{v} \quad \forall v, t
    \label{eq:vehicleElectricDrivingLimit}\\
    \begin{multlined}
    FS_{v} \left (D_{m,t} - \overline{Z}^{\text{EL}}_{v,t}\right) \phi_{v} \leq z^{\text{ICE}}_{v,t} \leq  \left (D_{m,t} - FS_{v} \, \overline{Z}^{\text{EL}}_{v,t}\right) \phi_{v}     \quad \forall v\in V_m, \forall m\in M^\text{ROAD}, \forall t
    \label{eq:vehicleIceDrivingLimit}
    \end{multlined}
\end{gather}

While vehicles with uncontrolled charging follow a predefined charging profile $X_{v,t}^\text{CON,IC}$, the other vehicles are able to choose the time for charging endogenously ($x_{v,t}^\text{CON}$). Endogenous charging is restricted in \Cref{eq:vehicleChargeing} considering a maximum flexible charging profile $\overbar{X}_{v,t}^\text{CON,FC}$.  
For the vehicles the state-of-charge of the battery $x^{\text{S}}_{v,t}$ has to lie within a minimum $\underbarnew{X}^{\text{S}}_{v,t}$ and a maximum state-of-charge profile $\overbar{X}^{\text{S}}_{v,t}$ in \Cref{eq:vehicleStateOfChargeLimits}.
The vehicle storage continuity in \Cref{eq:vehicleStateOfCharge} considers the charging efficiency $\eta^\text{CH}_{v}$.

\begin{gather}
    \begin{multlined}
        FS_v X_{v,t}^\text{CON,IC}  \phi_{v} \leq x_{v,t}^\text{CON} \leq FS_v X_{v,t}^\text{CON,IC}  \phi_{v} + (1-FS_v)  \overbar{X}_{v,t}^\text{CON,FC} \phi_{v} \quad \forall v, t
    \label{eq:vehicleChargeing}
    \end{multlined}\\    
    \begin{multlined}
    \left (1-FS_{v}\right) \underbarnew{X}^\text{S}_{v,t}\, \phi_{v} \leq x^{\text{S}}_{v,t}
    \leq  \left (1-FS_{v}\right) \overbar{X}^\text{S}_{v,t}\, \phi_{v} \quad \forall v, t
    \label{eq:vehicleStateOfChargeLimits}
    \end{multlined}\\
    \begin{multlined}
    x^{\text{S}}_{v,t+1}= x^{\text{S}}_{v,t} - EC_{v} \left (z^{\text{EL}}_{v,t} - FS_v \, \overline{Z}^{\text{EL}}_{v,t}  \, \phi_v\right)
    + \eta^{\text{CH}}_{v} \left( x_{v,t}^\text{CON} - FS_v \, X_{v,t}^\text{CON,IC} \, \phi_v \right) 
        \quad \forall v, t
    \label{eq:vehicleStateOfCharge}
    \end{multlined}
\end{gather}    

The energy demand is than calculated with the help of a distance-specific electricity consumption $EC_{v}$ and a distance-specific fuel consumption $FC_{v}$. The fuel consumption of vehicle units is described in \Cref{eq:fuelConsumptionVehicle}.  

\begin{gather}
    y^\text{CON}_{v,t} = FC_{v} \, z^{\text{ICE}}_{v,t} \quad \forall v,t
    \label{eq:fuelConsumptionVehicle}
\end{gather}

Variable internal combustion engine cost $C_{v}^\text{ICE}$ enter the objective function with $\sum_{v\in V} C_v^\text{ICE} y^\text{CON}_{v,t}$.

The possibility of power infeed in the grid (vehicle-to-grid) is not considered here.

\subsubsection{Cross-border electricity trade}

Power transfer $x_{i,j,t}$ between countries $i$ and $j \in I$ is modelled with a linear power transmission efficiency factor $TL_{i,j}<1$ where $x_{i,j,t} \leq NTC_{i,j,t}$ with predefined net transfer capacities ($NTC$).

\subsection{Electricity market clearing}

Proper operation of the electricity market in country $i \in I$ is ensured by the market clearing constraint in \Cref{eq:wholesaleElectricity}, which balances power generation and consumption of all participating technologies for each considered time step $t\in T$.

\begin{align}
\begin{multlined}[b]
    D_{m,t} + \sum_{\theta\in \Theta_m} x^\text{CON}_{\theta,t} + \sum_{j\in J_i} x_{i,j,t} 
    =\sum_{\gamma\in\Gamma_m} x^\text{GEN}_{\gamma,t} +\sum_{j\in J_i} TL_{j,i}\, x_{j,i,t} \quad \forall m\in M^\text{EL}_i, \forall i,t
\end{multlined}
\label{eq:wholesaleElectricity}
\end{align}

Hereby $\Gamma = G \cup R \cup U \cup A$ is the aggregate set of power producing units, $\Theta = G^\text{CHP} \cup H \cup U \cup A \cup B \cup V \cup O \cup L$ is the aggregate set of power consuming units and $J_i$ is the set of countries connected to country $i$.
Note that the dual variable of the market clearing constraint in \Cref{eq:wholesaleElectricity} can be extracted from the linear programming problem instance and interpreted as a proxy for the wholesale electricity market price, e.g. see \citep{Hartel.2021}.
Hereby we assume double-sided auctions and a uniform pricing regime with complete information of all market participants abstracting from strategic bidding behaviour.

\section{Case study description}
\label{S:Casestudy}

The analysed scenario for 2050 is based on a 95\,\% GHG reduction target in Europe \citep{Trafo_Bericht.2019}\footnote{More information about input data and results of this scenario is available on \href{https://openenergy-platform.org/}{openenergy-platform.org} in tables labelled \texttt{fh\_iee\_trafo\_fw} in the scheme \texttt{scenario} and the scenario name \texttt{fh\_iee\_trafo\_fw\_wenig\_dez\_bio\_mod\_sani}.}.
Therefore, energy-related GHG emissions have to be zero. The only possibility to use a storable fuel is to either produce methane with power-to-gas plants in Europe or to import renewable methane from sites outside of Europe for an assumed price of $P^{CH_4}$= 119.2\,EUR/MWh$\textsubscript{th}$~\citep{Pfennig.2017} which encompasses 5.2\,EUR/MWh$\textsubscript{th}$ for national transport. We do not consider power plants with carbon capture and storage (CCS).

The conventional power demand in Germany decreases to 456\,TWh\textsubscript{el} due to efficiency increases.
In the heating sector, we also model a considerable reduction of final energy demand as we assume an increase in the annual rate of building insulation to 1.7 \%/year for residential buildings and to 1.9 \%/year for non-residential buildings.

\par
The installed power generation and consumption capacities for the German market area are taken from the optimised European energy system scenario documented in \citep{Trafo_Bericht.2019}, see \Cref{Tab.instCap}.

\begin{table}[H]
\small
\centering
\begin{tabular}{l r r}
\hline
Technology & Installed capacity in GW$\textsubscript{el}$ \\
\hline
Solar PV & 174.4 \\
Onshore wind\footnotemark[3] & 162.3 \\ 
Offshore wind & 36.7 \\
Hydropower (incl. PHS) & 9.2 \\
CHP (OCGT + CCGT) & 40.3 \\
OCGT & not limited\\
Power-to-gas & 27.1 \\
Battery storage & 6.5 \\
\hline
\end{tabular}
\caption{Installed power generation and consumption capacities in the German market area (PHS - pumped-hydro storage), based on own calculations.}
\label{Tab.instCap}
\end{table}
\footnotetext[3]{Onshore wind power plants are a mix of plants with different rotor to generator ratio.}

Note that the only difference to this scenario is that OCGTs, as the most expensive power generation option, are not limited in size in our case study setup.
As for the stylised cases in the first part of the analysis that omit cross-border exchange with neighbouring countries and consider only the German market area by itself, this is to ensure that the electricity demand is met at any time.
In the second part of the analysis, the aggregated import capacity of Germany is assumed to equal 45\,GW\textsubscript{el} and the export capacity is 43\,GW\textsubscript{el}.
Transmission efficiency factors between bidding zones $TL_{i,j}$ vary between 84.8\,\% and 98.7\,\% depending on the line type and distance.

Variable operation costs and curtailment costs for variable RES are given in \Cref{Tab.EEvarCost} whereas the curtailment bonus $C^\text{CU}_{r}$ is used for better numerical performance as it ranks the curtailment of different technologies in surplus situations.
\begin{table}[H]
\footnotesize
\centering
\begin{tabular}{l r r}
    \hline
    $r$ & $C^\text{VP}_{r}$ in EUR/MWh$\textsubscript{el}$ & $C^\text{CU}_{r}$ in EUR/MWh$\textsubscript{el}$\\
    \hline
    Solar PV & 0.0 & 0.000009 \\
    Onshore wind & 4.6 & 0.000010 \\
    Offshore wind & 0.0 & 0.000008 \\
    \hline
\end{tabular}
\caption{Variable production cost and curtailment bonus for variable RES.}
\label{Tab.EEvarCost}
\end{table}

The round-trip efficiency of pump-hydro storage $\eta^\text{P}_u$ varies among the plants according to their age and turbines but averages about 73\,\%.

OCGTs are assumed to have an efficiency $\eta^\text{GEN}_{g}$ of 40\,\% and load change costs $C_g^\text{LC}$ of 4.8\,EUR/MWh\textsubscript{el}.

Power-to-gas plants have a power-to-fuel conversion factor $PF_{l}$ of 59.25\,\%, i.e. 75\,\% electrolysis and 79\,\% methanation reactor, as well as load change costs $C_l^\text{LC}$ of 1\,EUR/MW\textsubscript{el}.
The opportunity costs for domestic methane production instead of importing energy carriers from outside of Europe is assumed to be 114\,EUR/MWh\textsubscript{th} ($P^{CH_4}$ minus national transport costs).

Lithium-ion batteries have a round-trip efficiency of $\eta^\text{IN}_{a}\cdot \eta^\text{OUT}_{a}=92\,\%$ and small linear storage losses $\lambda_{a}^\text{S}$ of 0.0000139\,\% per hour.
The electricity storage self-discharge factor $\Lambda_{a}^\text{S}$ is set to zero, and the storage capacity ratio corresponds to 6\,MWh\textsubscript{el}/MW\textsubscript{el}.
Its ramping cost are set to 0.1\,EUR/MW\textsubscript{el}.

Air conditioning is modelled with an installed capacity of 21.1\,GW$\textsubscript{el}$ and 12\,hour storage.
We assume linear storage losses $\lambda^{\text{S}}_{o}$ of 10\,\% per hour whereas $\Lambda^{\text{S}}_{o}$ is set to zero.

Further assumptions for CHP units in hybrid CHP systems are given in \Cref{Tab.CHPsystems}. These include parameters for the electrical (condensing mode) efficiency ($\eta_\text{el}$), power-to-heat ratio (backpressure limit) factor ($PH_{g}$), and power loss factor ($PL_{g}$) of the CHP plant, as well as the design ratio factor of maximum thermal CHP output in relation to the maximum heat demand ($\Pi^\text{CHP}_{g}$), and the thermal efficiency of the backup condensing boiler ($\eta^\text{CB}_{g}$).

\begin{table}[H]
\small
\centering
\begin{tabular}{l r r r r r}
    \hline
    System type & $\eta_\text{el}$ & $PH_{g}$ & $PL_{g}$ & $\Pi^\text{CHP}_{g}$ & $\eta^\text{CB}_{g}$\\
    \hline
    OCGT CHP & 0.42 & 0.86 & 0.01 & 0.33 & 0.90 \\
    Small CHP & 0.46 & 1.10 & 0.00 & 0.20 & 0.93\\
    CCGT CHP & 0.56 & 1.40 & 0.10 & 0.33 & 0.93 \\ 
    \hline
\end{tabular}
\caption{CHP system types and their characteristic parameters, own assumptions.}
\label{Tab.CHPsystems}
\end{table}

The design factor of the fuel boiler system $\Pi^\text{BU}_g$ equals 2 so that heat demand peaks can always be supplied whereas the design factor of the backup electrical heaters $\Pi^\text{CON}_g$ is lower with 0.8-1.9.

Load change costs of the CHP plant $C_g^\text{LC}$ equal 4.8\,EUR/MWh\textsubscript{el}.
Hybrid CHP systems further consist of either an electric boiler with an efficiency of 0.99\,MWh\textsubscript{th}/MWh\textsubscript{el} or a large-scale heat pump with COPs between 3.3 and 4\,MWh\textsubscript{th}/MWh\textsubscript{el}.
Losses in thermal storage systems are described by a linear storage loss coefficient of 0.21\,\%/h.

The thermal storage has an efficiency of $\eta^{\text{IN}} = \eta^{\text{OUT}} = 99\,\%$ and is sized between 1 and 4\,MWh\textsubscript{th}/MW\textsubscript{th} for industrial applications and 8\,MWh\textsubscript{th}/MW\textsubscript{th} for district heating purposes.

Hybrid boiler systems consist of fuel boilers with an efficiency ranging from 0.9 to 0.93\,MWh\textsubscript{th}/MWh\textsubscript{fuel} and electric boilers with an efficiency of 0.99\,MWh\textsubscript{th}/MWh\textsubscript{el}.

For heat pump systems, the heat pump units feature an hourly predefined COP while the efficiency of the backup unit equals 0.90\,MWh\textsubscript{th}/MWh\textsubscript{fuel} (methane) or 0.99\,MWh\textsubscript{th}/MWh\textsubscript{el} (power), see also \citep{Hartel.2020b}.
Thermal storage energy-to-power capacity ratios correspond to 6\,MWh\textsubscript{th}/MW\textsubscript{th} for space heating and 4\,MWh\textsubscript{th}/MW\textsubscript{th} for hot water heating purposes.

\Cref{Tab.vehicles} shows the composition of the vehicle fleet in 2050. 
While 80\,\% of all BEV are charged under a flexible charging behaviour regime, only 60\,\% of PHEVs feature controlled charging behaviour ($FS_{v}$).
Depending on the vehicle size, the efficiency of the electric engine of BEV and hybrid cars ranges from 0.14 to 0.25\,kWh\textsubscript{el}/km.
The round-trip efficiency of the vehicle battery is assumed to equal 90\,\%.

\begin{table}[H]
\small
\centering
\begin{tabular}{l r r}
\hline
Technology & Passenger cars & Trucks\\
 & (mio. vehicles) & (mio. vehicles)\\
\hline
    BEV & 14.3 & 0.3 \\
    Hybrid & 10.7 & 0.1\\
    Fuel (PtL, H\textsubscript{2}) & 4.1 & 0.3 \\
    Overhead-line & - & 0.3\\
\hline
\end{tabular}
\caption{Number of vehicles per technology type in Germany (PtL - Power-to-Liquid-fuel, H\textsubscript{2} - hydrogen), own assumption based on own computations \citep{Trafo_Bericht.2019,Boettger.2018,Trost.2017}.}
\label{Tab.vehicles}
\end{table}

All time series used in the model are based on the weather year and calendar year 2012.

\section{Results}
\label{S:Results}

Note that the power price time series presented in this paper are available in an open-access repository at 10.5281/zenodo.4613964.

\subsection{Power price setting of single technologies}

To understand the price setting of the individual technologies participating in future low-carbon wholesale electricity markets, we start the analysis with a very simplified power system with one bidding zone, i.e. Germany.
In the following, we put wind and solar PV in the system and gradually add generation and consumption technologies.
Note that in the following we always put the complete RES generation in the system although the power demand is much lower in the single cases as we always neglect a considerable part of the sector coupling technologies. This results in a high number of hours in our analyzed cases year where RES have to be curtailed.

\subsubsection{Gas turbines and RES}

\begin{figure}[h!tb]
    \centering\includegraphics[width=\linewidth/2]{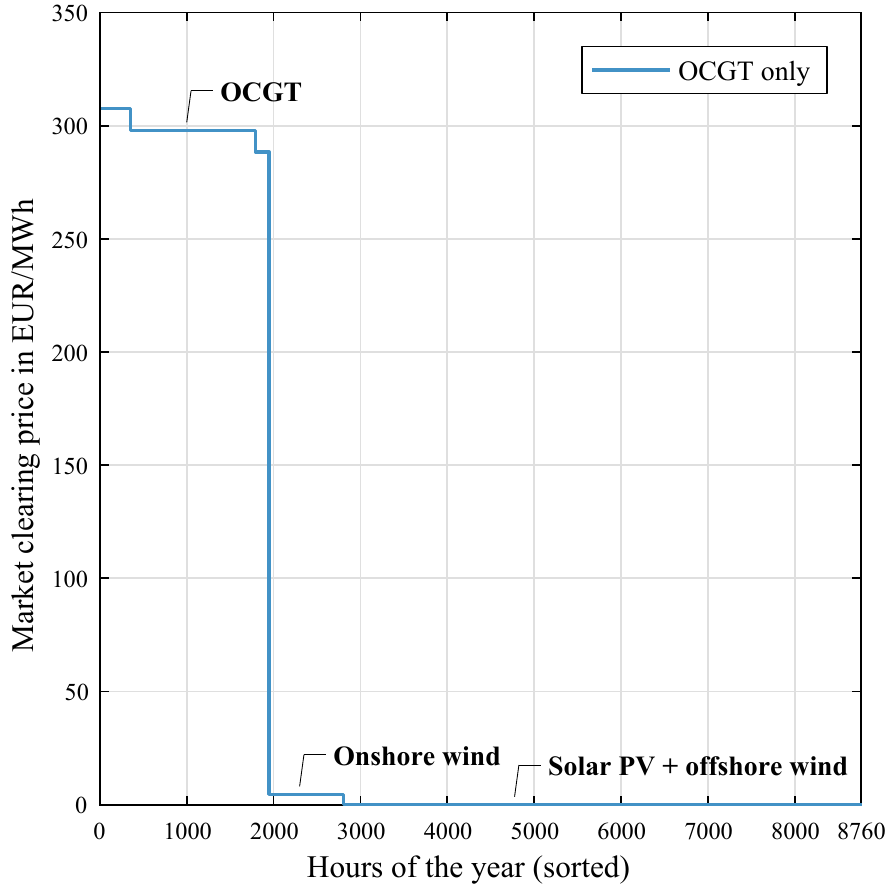}
    \caption{Market clearing price duration curve in a system with RES and OCGTs, own illustration based on own computations.}
    \label{Fig.Prices_EE_GT}
\end{figure}

Fig.~\ref{Fig.Prices_EE_GT} shows the market clearing price duration curve in a system with RES and OCGTs.
In this very simplified system, there are generally two types of market outcomes.   
On the one hand, there are situations with sufficient RES generation to meet the electricity demand, resulting in very low power prices.
On the other hand, there are times where the power demand exceeds RES generation and OCGTs supply the residual demand at their comparably high marginal costs.
A closer look at the power price curve exhibits that OCGTs generally set a power price of 298\,EUR/MWh\textsubscript{el}, which equals the renewable methane import price of 119.2\,EUR/MWh$\textsubscript{fuel}$ divided by the gas turbines efficiency $\eta^\text{GEN}_{g} = 0.4$.
Furthermore, depending on the required load change decisions, the gas turbine also sets prices of $298\pm2\cdot C_g^\text{LC}$ of 4.8\,EUR/MWh$\textsubscript{el}$, i.e. 307.6 and 288.4\,EUR/MWh\textsubscript{el}.
When OCGT generation increases in one hour and decreases in the next, or vice versa, the bids incur the ramping costs twice.
Conversely, when the OCGT power output is continuously increasing or decreasing, the ramping costs cancel out in the shadow price formulation.
Moreover, there is a price duration curve step that corresponds to the marginal cost of onshore wind $C^\text{VP}_{Onshore Wind}= 4.58$~EUR/MWh$\textsubscript{el}$, also indicating when onshore wind is curtailed.

\subsubsection{Gas turbines, RES and Power-to-Gas}

\begin{figure}[h!tb]
    \centering\includegraphics[width=\linewidth/2]{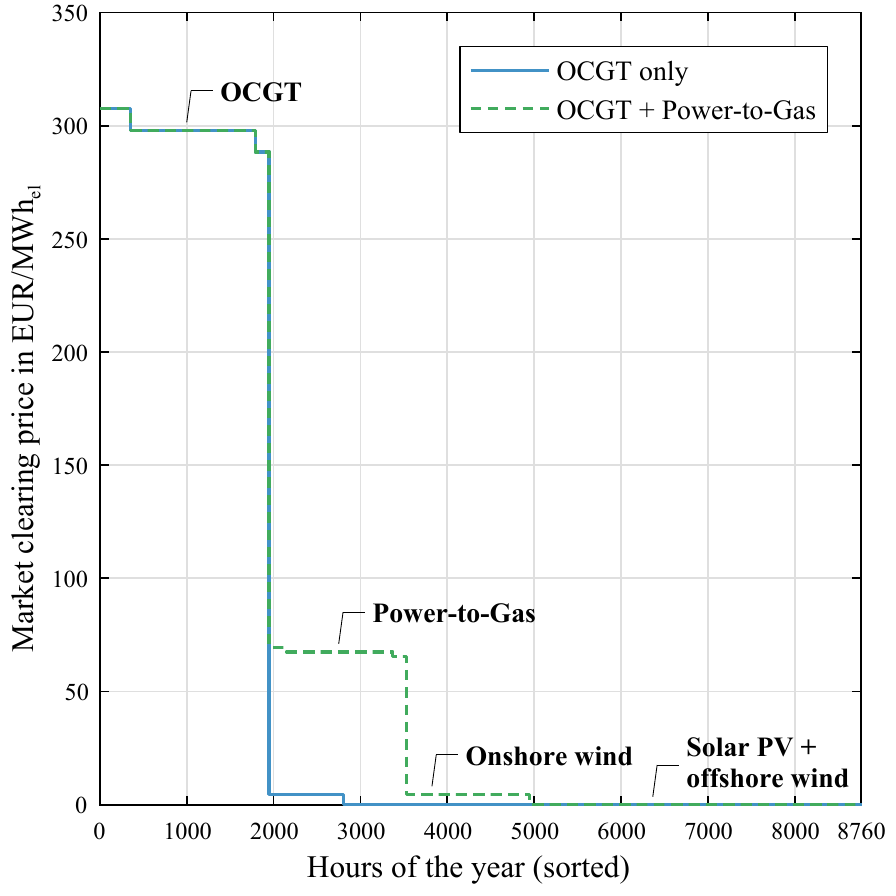}
    \caption{Market clearing price duration curve in a system with RESs, OCGTs and power-to-gas plants, own illustration based on own computations.}
    \label{Fig.Prices_EE_GT_P2G}
\end{figure}

Fig.~\ref{Fig.Prices_EE_GT_P2G} depicts the market clearing price duration curve for the system extended by a power-to-gas plant.
The power-to-gas plant is an additional power consumer that uses excess RES generation to produce methane.
For the first 2,000 hours, during which the OCGT is setting the prices, the price duration curve in Fig.~\ref{Fig.Prices_EE_GT_P2G} is identical to the one in Fig.~\ref{Fig.Prices_EE_GT}.
The power-to-gas plant itself has opportunity costs of 67.55\,EUR/MWh$\textsubscript{el}$ resulting from 114\,EUR/MWh$\textsubscript{fuel}$ gas opportunity costs (renewable methane-import price of 119.2\,EUR/MWh\textsubscript{el} minus national transport costs of 5.2\,EUR/MWh) times the unit's efficiency (59.25\,\%), see also \citep{Ruhnau.2020}.
Furthermore, there are small price steps around the 67.55\,EUR/MWh$\textsubscript{el}$ resulting from load change costs $C_l^\text{LC}$ of 1\,EUR/MW\textsubscript{el}.
With the same explanation already given for the OCGT units, the ramping costs enter twice in the shadow price.
Due to the power-to-gas plant's predefined capacity, it sets the price in approximately 1,500\,hours of the considered year.
In the remaining hours, the RES production exceeds the power-to-gas capacity and needs to be curtailed.

\subsection{Effect of storage technologies}

Technologies that are able to shift generation and consumption over time, such as battery storage in the power sector or thermal storage in the heating sector, have a smoothing effect on the shape of the power price curve.
We analyse in detail the effects of battery storage, (pumped) hydro storage, and thermal storage in the following subsections.

\subsubsection{Battery storage}

\begin{figure}[htb]
    \centering\includegraphics[width=\linewidth/2]{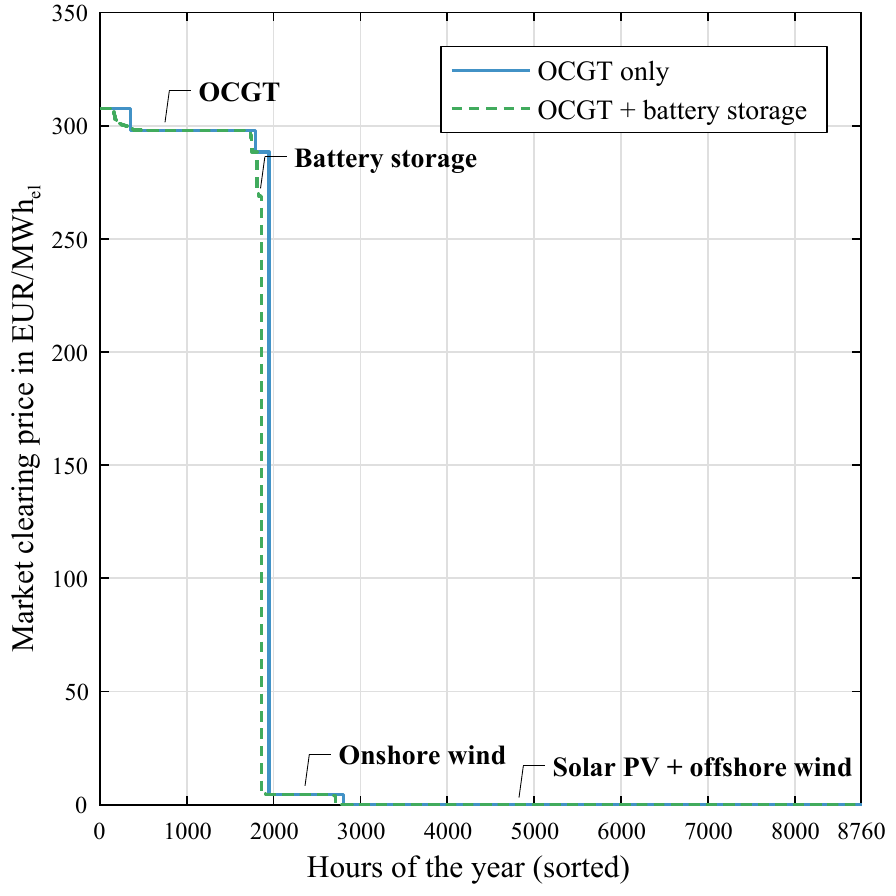}
    \caption{Market clearing price duration curve in a system with RESs, OCGTs and battery storage systems, own illustration based on own computations.}
    \label{Fig.Prices_EE_GT_Batterie}
\end{figure}

As can be seen from Fig.~\ref{Fig.Prices_EE_GT_Batterie}, battery storage plants reduce the dispatch hours of the OCGTs as they are able to shift RES generation in the temporal domain.
Furthermore, the battery storage systems cause a smoothing of OCGT price duration steps.
Additionally, there are a few hours during which battery storage systems are setting the electricity price with opportunity costs of 268.40\,EUR/MWh\textsubscript{el}, which equals the marginal costs of OCGTs times the round-trip efficiency of the battery systems.

\subsubsection{Conventional and pumped hydropower}

\begin{figure}[h!tb]
    \centering\includegraphics[width=\linewidth/2]{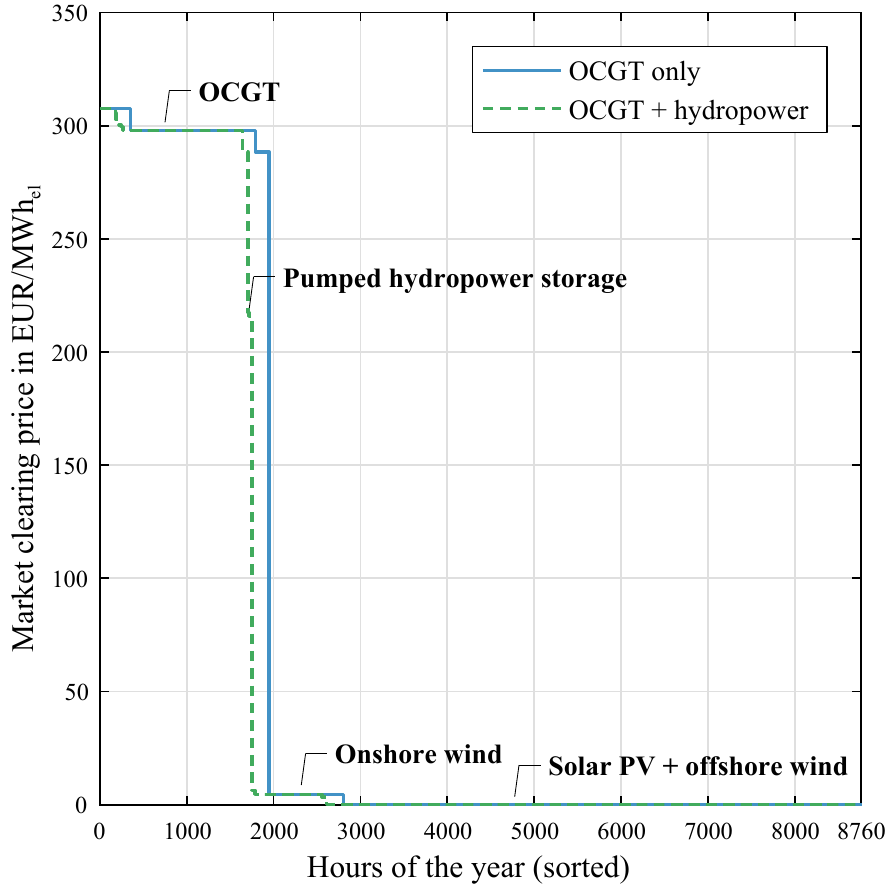}
    \caption{Market clearing price duration curve in a system with RESs, OCGTs, and hydropower systems, own illustration based on own computations.}
    \label{Fig.Prices_EE_GT_Hydro}
\end{figure}

In general, the effect of pumped hydropower is comparable to battery storage with the difference that hydropower includes natural inflow in the upstream reservoirs which reduces the operating hours of OCGTs, see Fig.~\ref{Fig.Prices_EE_GT_Hydro}.
Nevertheless, the smoothing effect on OCGT price steps are similar and pumped hydropower sets prices in a range of 217\,EUR/MWh\textsubscript{el}, which equals OCGT marginal costs times the round-trip efficiency of the pumped hydro plants (approximately 73\,\%).

\subsubsection{Electric vehicles with batteries}

\begin{figure}[h!tb]
    \centering\includegraphics[width=\linewidth/2]{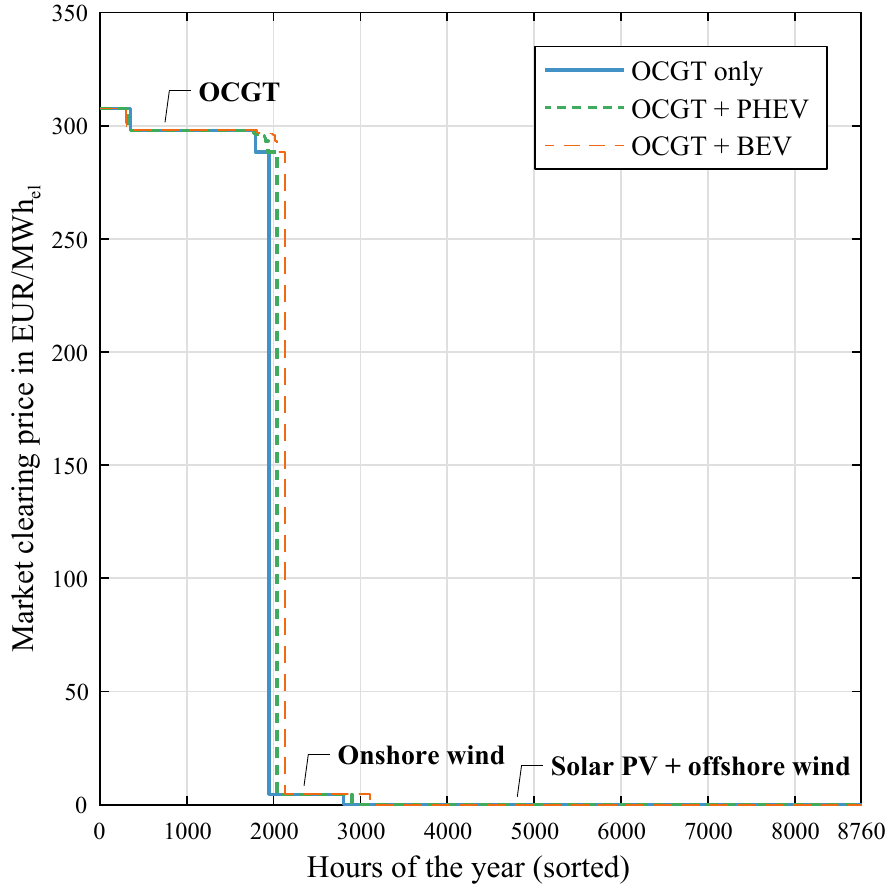}
    \caption{Market clearing price duration curve in a system with RESs, OCGTs, and electric vehicles, own illustration based on own computations.}
    \label{Fig.Prices_EE_GT_Vehicles}
\end{figure}

Fig.~\ref{Fig.Prices_EE_GT_Vehicles} shows the effects of hybrid and pure electric vehicles on the the market clearing and the formation of wholesale electricity prices.
First, the additional electricity demand of the transport sector shifts the price curve to the right, increasing the operating hours of OCGTs in the stylised system configuration.
Second, the flexible charging behavior of battery-electric and plug-in electric vehicles results in a partial smoothing of the price steps observed for the OCGT units.

PHEV have opportunity costs of 311.89\,EUR/MWh\textsubscript{el} in our scenario setting (fuel price / efficiency of internal combustion engine / efficiency of electric engine) which are higher than the marginal costs of the most expensive generation technology. This means that in times with power prices higher than the marginal costs of PHEV they would not charge and use the internal combustion engine instead. As this is never the case in our setting, we do not observe this price step.

\subsubsection{Thermal storage}

\begin{figure}[h!tb]
    \centering\includegraphics[width=\linewidth/2]{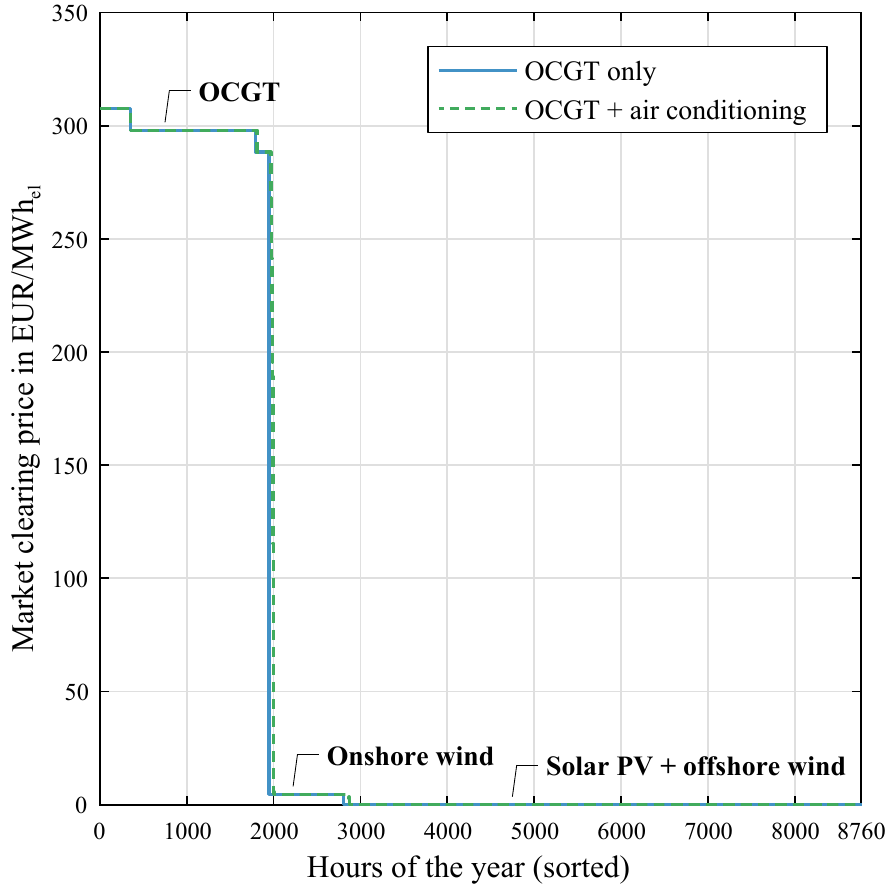}
    \caption{Market clearing price duration curve in a system with RESs, OCGTs, and thermal storage for air conditioning systems, own illustration based on own computations.}
    \label{Fig.Prices_EE_GT_Klimatisierung}
\end{figure}

The model includes the demand for air conditioning and the modelling approach includes a thermal storage for air conditioning, see \Cref{eq:thermalStorageContinuityAirCond}.
As can be seen in Fig.~\ref{Fig.Prices_EE_GT_Klimatisierung}, the introduction of additional electricity demand for air conditioning shifts the power price curve slightly to the right while the effect on the price setting is rather small. Due to the thermal storage availability, there are a few hours with prices between the marginal costs of OCGTs and the variable production cost of onshore wind.
Even fewer hours exhibit market clearing prices between the variable production cost of onshore wind and zero.

\subsection{Effect of hybrid heating supply technologies}

Next, we analyse and highlight the important effects of different hybrid heat supply technology combinations on the wholesale electricity market prices.

\subsubsection{Heat pump systems with thermal storage}

\begin{figure}[h!tb]
    \centering\includegraphics[width=\linewidth/2]{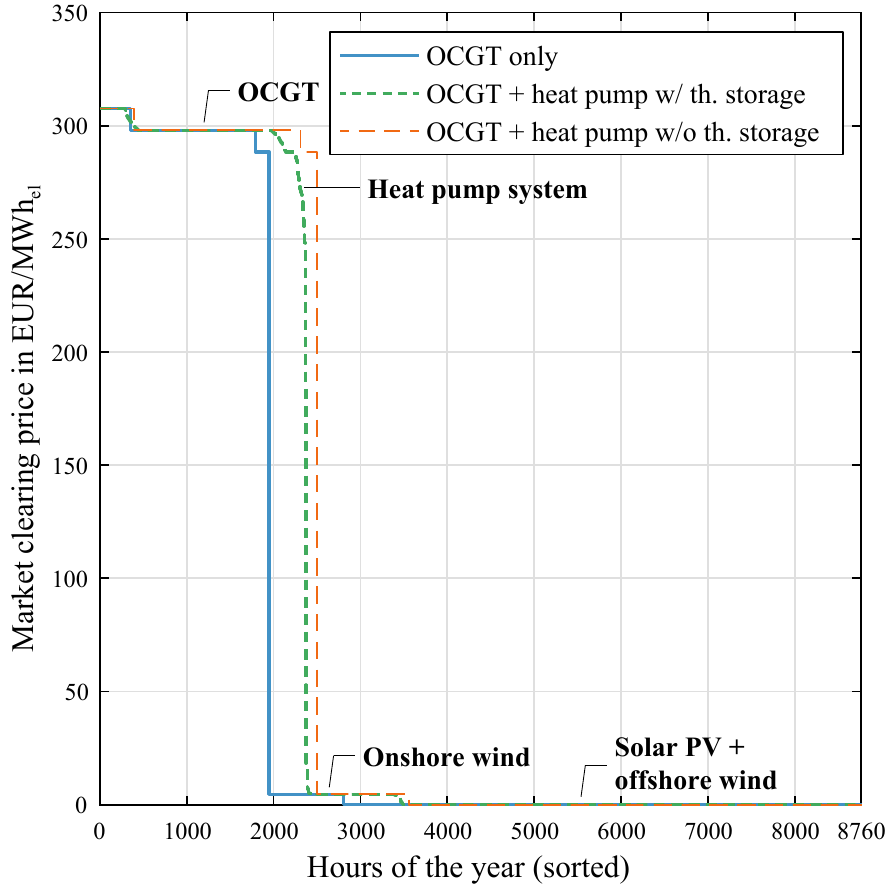}
    \caption{Market clearing price duration curve in a system with RESs, OCGTs, and heat pump systems, own illustration based on own computations.}
    \label{Fig.Prices_EE_GT_WP}
\end{figure}

Fig.~\ref{Fig.Prices_EE_GT_WP} displays the price formation effects of decentralised heat pumps.
A first observation is that the additional power demand shifts the price curve to the right.
If there were no thermal storage available, no additional steps would occur in the price duration curve.
However, given the availability of thermal storage, new price steps become visible due to the opportunity arising from either the immediate or temporally delayed use of electricity to supply final heating demands.

\subsubsection{Hybrid systems with electric and fuel boiler units}

\begin{figure}[h!tb]
    \centering\includegraphics[width=\linewidth/2]{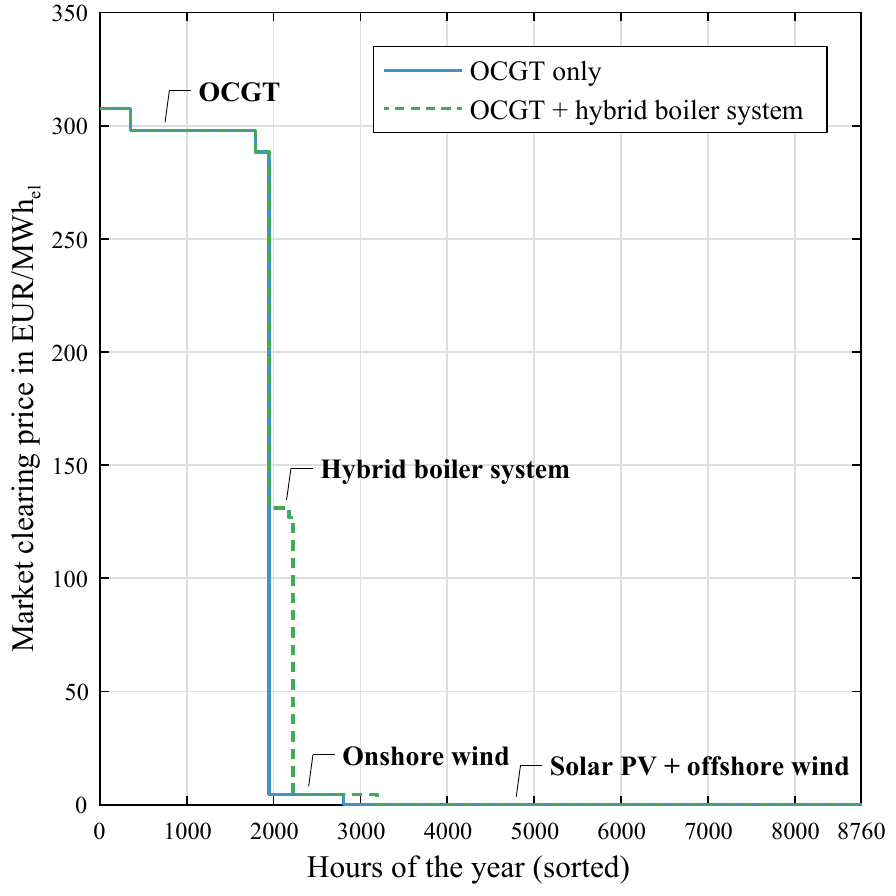}
    \caption{Market clearing price duration curve in a system with RESs, OCGTs, and hybrid boiler systems, own illustration based on own computations.}
    \label{Fig.Prices_EE_GT_Boiler}
\end{figure}

Next, we focus on hybrid heat supply systems consisting of condensing boilers and electric boilers in Fig.~\ref{Fig.Prices_EE_GT_Boiler}.
Due to the choice of using gas boilers (marginal cost = gas prices/efficiency) and electric boilers, there are two new price steps in the price duration curve, i.e. 131.20 and 126.89\,EUR/MWh\textsubscript{el}, respectively, depending on the fuel boilers efficiency $\eta^\text{CB}_{b}$ between 90\,\% and 93\,\%.
These marginal values can be reconstructed using the formula $\frac{P^{CH_4}}{\eta^\text{CB}_{b}}\eta^\text{CON}_b$.

\subsubsection{Hybrid systems with CHP units}

The calculation of opportunity costs for hybrid CHP systems $g \in G^\text{CHP}$ consisting of a CHP plant, a fuel boiler, thermal storage, and an electric heating option, i.e. electric boiler or heat pump unit, is more complex.
First, for the CHP units, the marginal costs depend on the ratio of fuel price to electric efficiency.
Depending on the hourly load change, there also exist price steps slightly increased or decreased by the ramping costs. 
Second, there is an additional opportunity when it comes to the supply of heat by using the CHP or the fuel boiler unit.
Therefore, in situations where fuel boilers could supply the heat demand instead of the CHP plant, the opportunity costs\footnote{See Table~\ref{Tab.OppcostBoiler} in \ref{Appendix.tables} for more details of the calculation.} can be calculated by 
\begin{equation}
        \frac{P^{CH_4}}{\eta^\text{GEN}_g} - \frac{P^{CH_4}}{\eta^\text{CB}_{g}\cdot PH_{g}} + P^{CH_4} \cdot \frac{ PL_{g}}{PH_{g}} \quad \forall g\in G^\text{CHP}\,,
    \label{eq:cost_Boiler_marginal}
\end{equation}
which is also dependent on the load change costs that have to be adapted to 
\begin{equation}
    (+ \text{ or } -) \quad 2 \cdot C_g^\text{LC} \cdot \Biggl(1+\frac{PL_{g}}{PH_{g}}\Biggr)\quad \forall g\in G^\text{CHP}\,,
    \label{eq:cost_Boiler_ramping}
\end{equation}
due to the interaction between the CHP and fuel boiler units.
Third, there is another opportunity between the heat extraction from the CHP plant and the electric boiler or heat pump unit.
The corresponding opportunity costs\footnote{See Table \ref{Tab.OppcostP2H} in \ref{Appendix.tables} for more details of the calculation.} can be calculated by
\begin{equation}
    \frac{\frac{P^{CH_4}}{\eta^\text{GEN}_g}}{1+\frac{1}{PH_{g}\cdot \eta^\text{CON}_g}} + \frac{P^{CH_4}\cdot \frac{PL_{g}}{PH_{g}}}{1+\frac{1}{PH_{g}\cdot\eta^\text{CON}_g}}\quad \forall g\in G^\text{CHP}\,,
    \label{eq:cost_P2H_marginal}
\end{equation}
and the ramping costs become
\begin{equation}
    (+ \text{ or } -) \quad\frac{2 \cdot C_g^\text{LC}}{1+\frac{1}{PH_{g}\cdot \eta^\text{CON}_g}} \cdot \Biggl(1+\frac{PL_{g}}{PH_{g}}\Biggr)\quad \forall g\in G^\text{CHP}\,.
    \label{eq:cost_P2H_ramping}
\end{equation} 

The idea behind this opportunity is that if the CHP plant is operated on the power-to-heat-ratio line in its feasible operating region, an increase in electric power generation leads to an equivalent increase in heat generation relative to the power-to-heat ratio.
In this case, the heat generation of the electric boiler can be reduced by this amount as the heat demand is inelastic and has to be supplied.
Consequently, the power demand of the electric boiler is reduced dependent on its efficiency $\eta^\text{CON}_g$.

\begin{figure}[h!tb]
    \centering\includegraphics[width=\linewidth/2]{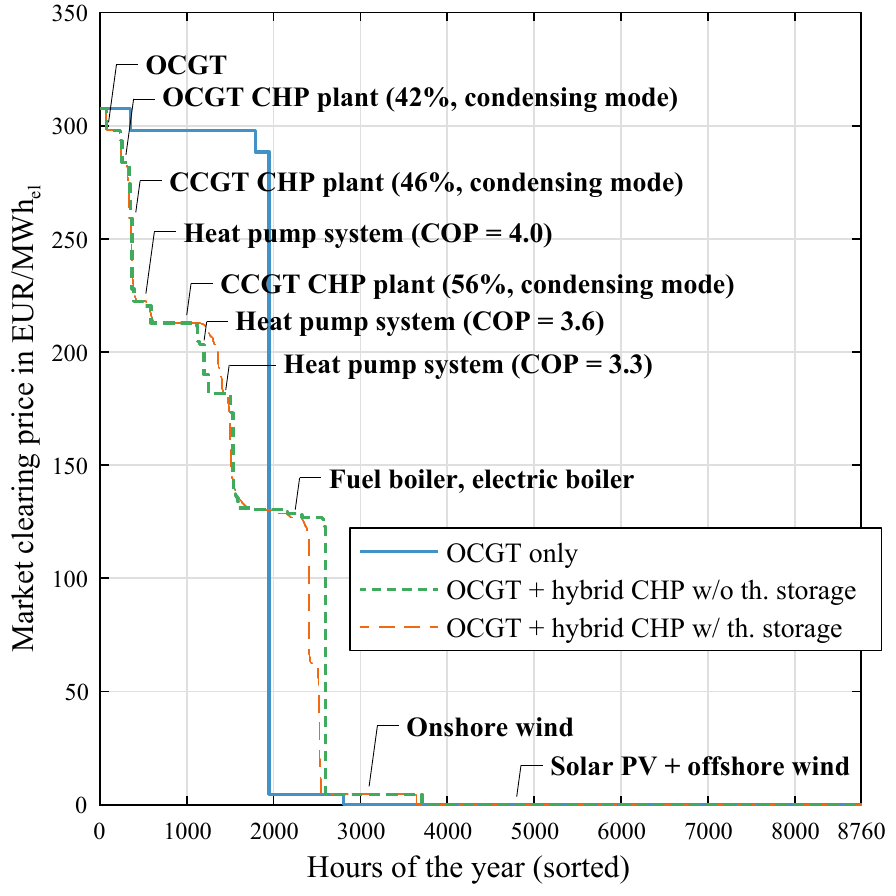}
    \caption{Price duration curve in a system with RESs, OCGTs, and hybrid CHP systems, own illustration based on own computations.}
    \label{Fig.Prices_EE_GT_chp}
\end{figure}

Fig.~\ref{Fig.Prices_EE_GT_chp} shows the market clearing price duration curve for the case with CHP systems with and without thermal storage. 
According to the assumed electrical efficiencies ranging from 42 to 56\,\% (condensing mode), CHP plants fueled by fuel (renewable methane) offer power with lower marginal costs (212 to 284\,EUR/MWh\textsubscript{el}) than OCGTs (298\,EUR/MWh\textsubscript{el}).
Therefore, CHP plants partly displace the OCGT, and there are considerably fewer hours in which OCGTs set the price.
Depending on the efficiencies for the CHP plant and the fuel boiler, there are price steps between 129 and 136\,EUR/MWh\textsubscript{el} where there is an opportunity between the CHP and fuel boiler units.
For electric boilers with a conversion efficiency of 99\,\%, there are price steps between 128 and 130\,EUR/MWh\textsubscript{el}.
Similarly, for heat pumps with COPs ranging from 3.3 to 4\,MWh\textsubscript{th}/MWh\textsubscript{el} the corresponding price steps are between 182 and 220\,EUR/MWh\textsubscript{el}.
Moreover, there are price steps at the corresponding marginal costs plus/minus the load change costs.
Regarding the effects of the included thermal storage systems, it stands out that, in the price range from 180 to 220\,EUR/MWh\textsubscript{el}, heat pump units are able to store their heat in the thermal storage.
The stored heat then replaces heat from fuel boilers at a price of about 130\,EUR/MWh\textsubscript{el}.

\begin{figure}[h!tb]
    \centering\includegraphics[width=\linewidth/2]{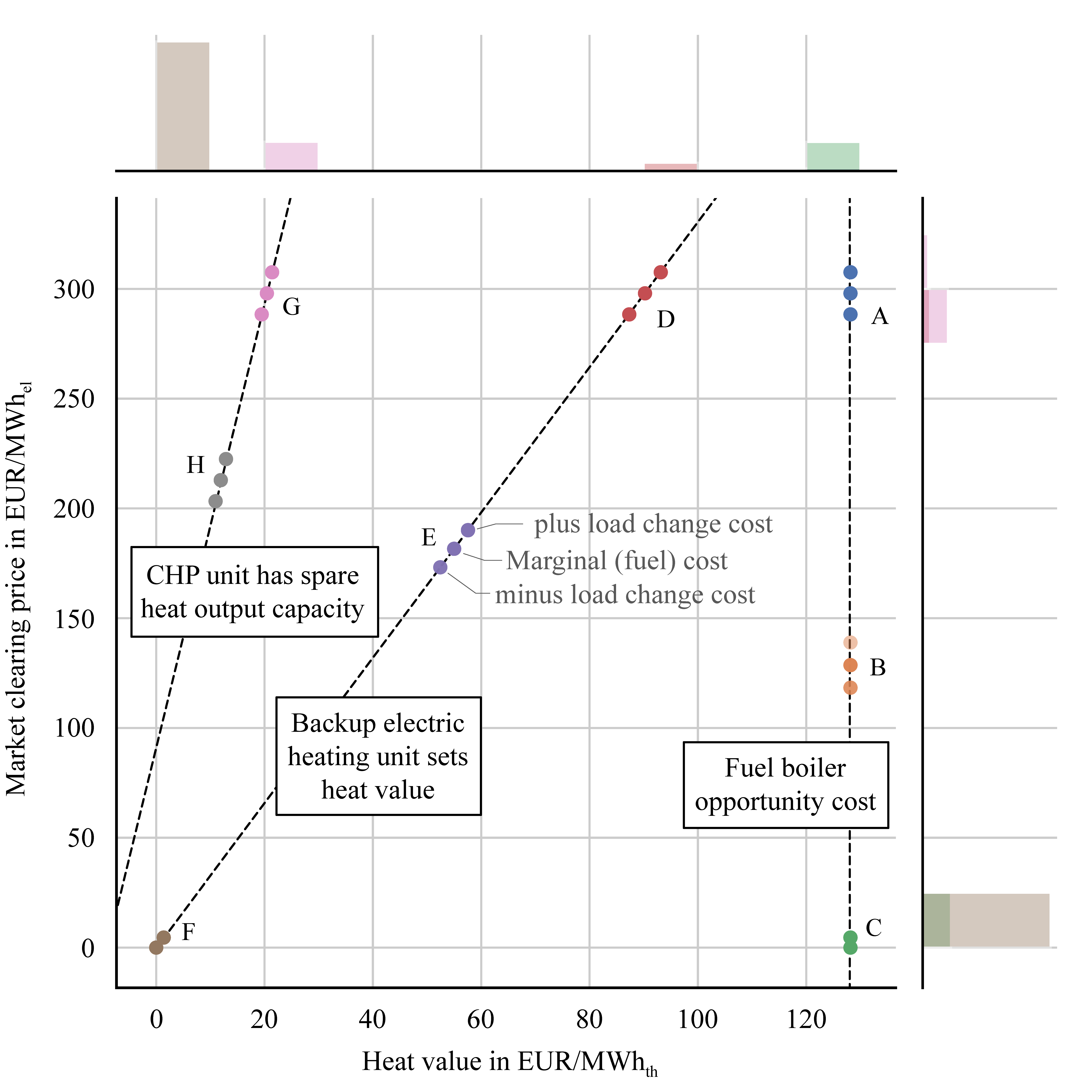} 
    \caption{Market clearing price versus the heat value of an exemplary multivalent CHP system, own illustration based on own computations. Bars at the figure border represent the frequency of the different states.}
    \label{Fig.chp_system_value_red}
\end{figure}

Fig.~\ref{Fig.chp_system_value_red} displays the dispatch of an exemplary multivalent CHP system against the electricity market clearing price and the corresponding heat value of each hour of the considered year.
The operation of the multivalent CHP system falls into eight distinct states, which are summarised below.
For a detailed elaboration on the the price formation of power and heat output in multivalent CHP systems, see Table \ref{TabCHPcases} in \ref{Appendix.tables}.

\begin{itemize}
    \item Heat value determined by backup fuel boilers (States A-C)
            \begin{enumerate}[label=\Alph*]
                \item Power price is set by OCGTs (123\,h)
                \item Opportunity between backup fuel boiler and CHP unit (10\,h)
                \item Excess power available and backup electric heating unit at full capacity (1,232\,h)
            \end{enumerate}
    \item Backup electric heating unit sets heating value depending on current electricity price (States D-F)
            \begin{enumerate}[label=\Alph*]
            \setcounter{enumi}{3}
                \item Power price is set by OCGTs (377\,h)
                \item Backup electric heating unit sets both the electricity price and heat value (49\,h)
                \item Excess power available and backup electric heating unit at partial load (5,565\,h)
            \end{enumerate}
    \item CHP unit has spare capacity for additional heat output (States G-H)
            \begin{enumerate}[label=\Alph*]
            \setcounter{enumi}{6}
                \item CHP unit at maximum boiler load (1,341\,h)
                \item CHP unit at partial boiler load with spare capacity for both power and heat production (63\,h)
            \end{enumerate}
\end{itemize}

\begin{figure}[h!tb]
    \centering\includegraphics[width=\linewidth/2]{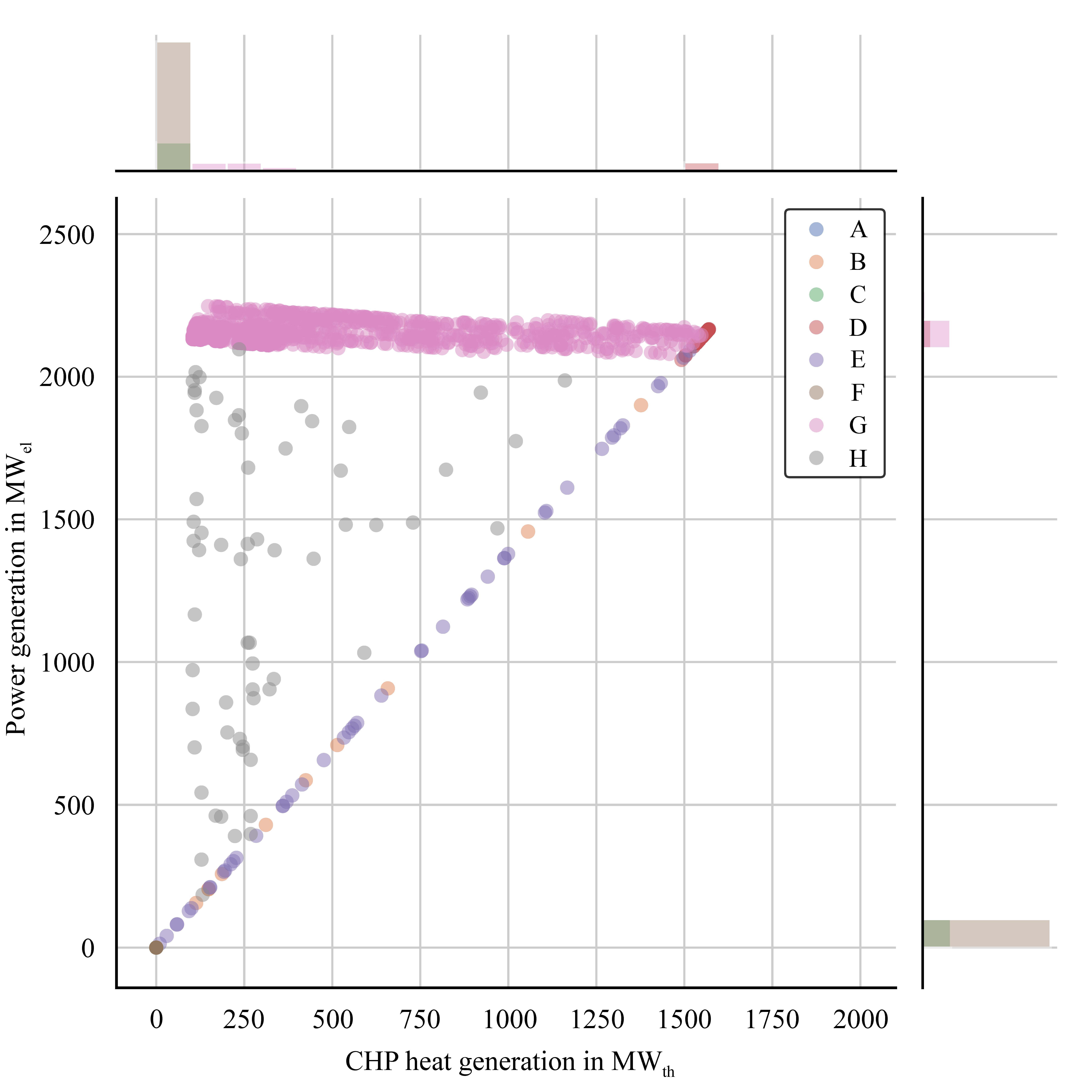}
    \caption{Power and CHP heat generation of an exemplary multivalent CHP system, own illustration based on own computations. Note that overlapping transparent points appear darker. Bars at the figure border represent the frequency of the different states.}
    \label{Fig.chp_system_pq_red}
\end{figure}

Fig.~\ref{Fig.chp_system_pq_red} displays the dispatch results (power and heat output) of the same exemplary cogeneration unit of a multivalent CHP system for each hour of the analysed scenario year. Note that we aggregate single CHP blocks to larger blocks for numerical reasons.
The identified operation states are again marked with the corresponding letters.
Note that due to the representation as a technology cluster, using a continuous linear model, there is no minimum stable generation limit visible here.

\subsection{Complete power price curve of all technologies}

\begin{figure}[h!tb]
    \centering\includegraphics[width=\linewidth/2]{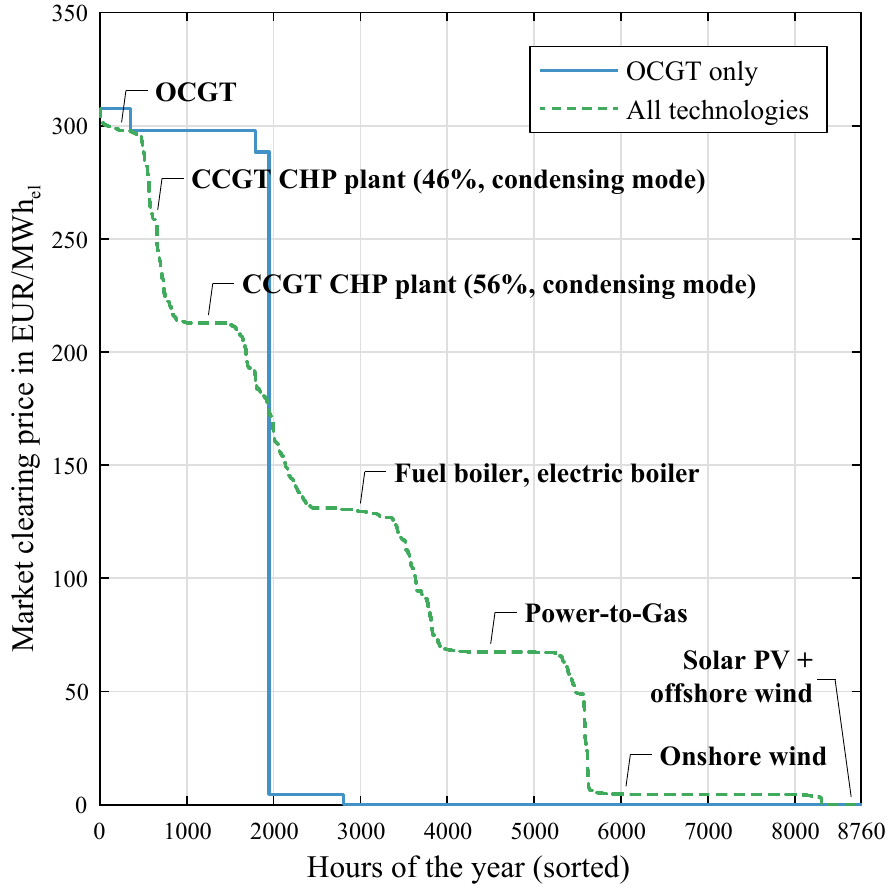}
    \caption{Market clearing price duration curve in the German system with all considered technologies, own illustration based on own computations.}
    \label{Fig.Prices_all}
\end{figure}

Fig.~\ref{Fig.Prices_all} displays the complete market clearing price duration curve for a system with all aforementioned technologies.
It becomes clear that it is much more heterogeneous than the price curve for the simple system purely based on RES and OCGT systems, recall Fig.~\ref{Fig.Prices_EE_GT}.
The longest price duration steps result from the OCGTs, CHP plants, fuel and electric boilers, and power-to-gas plants.
The curve between these steps is comparably smooth due to the combined effect of multiple storage technologies.
Under the assumption of a double-sided auction and a uniform pricing regime, either the available and dispatched generation technology with the highest marginal cost or the power consumption technology with the lowest marginal value sets the power price.

\subsection{Cross-border integration effects}

After the analysis of the power price setting of the single technologies, we focus on cross-border integration effects which are also crucial for the price formation in low-carbon electricity markets, see \citep{Hartel.2021}. 
In a first step, we increase the spatial scope of the analysis previously focused on Germany by one (large) neighboring country (France) with a net transfer capacity of 4.8 GW\textsubscript{el}.

\begin{figure}[h!tb]
    \centering\includegraphics[width=\linewidth/2]{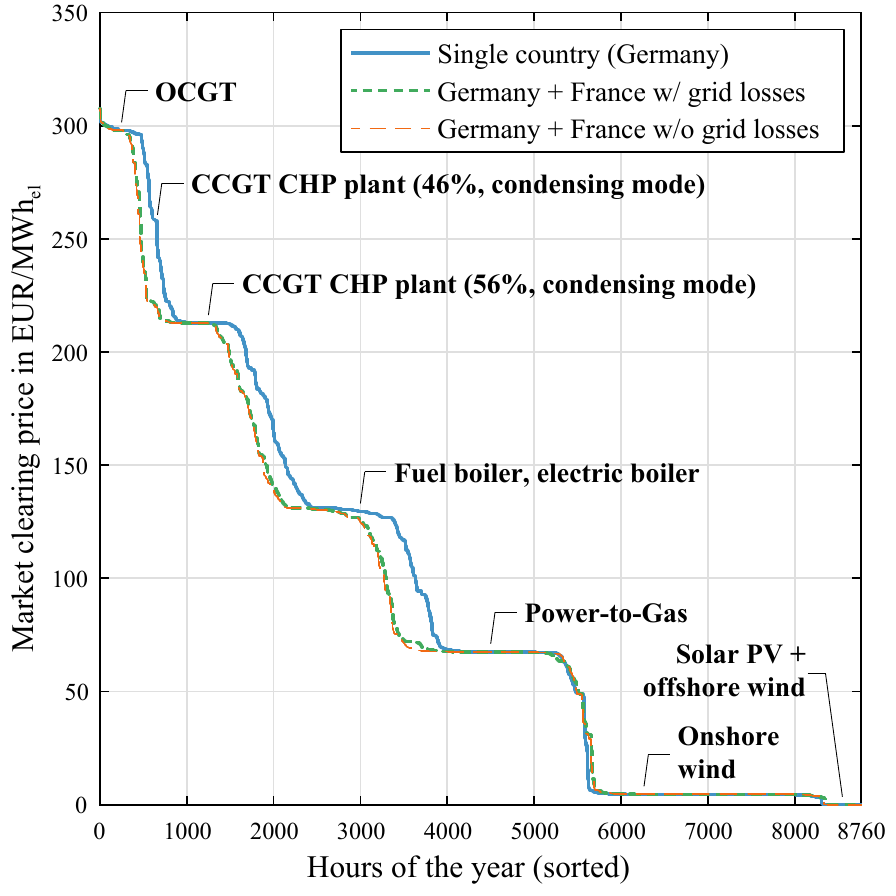}
    \caption{Market clearing price duration curve in the German system with all considered technologies, own illustration based on own computations.}
    \label{Fig:price_duration_comparison_deu_fra}
\end{figure}

\begin{figure}[h!tb]
    \centering\includegraphics[width=\linewidth/2]{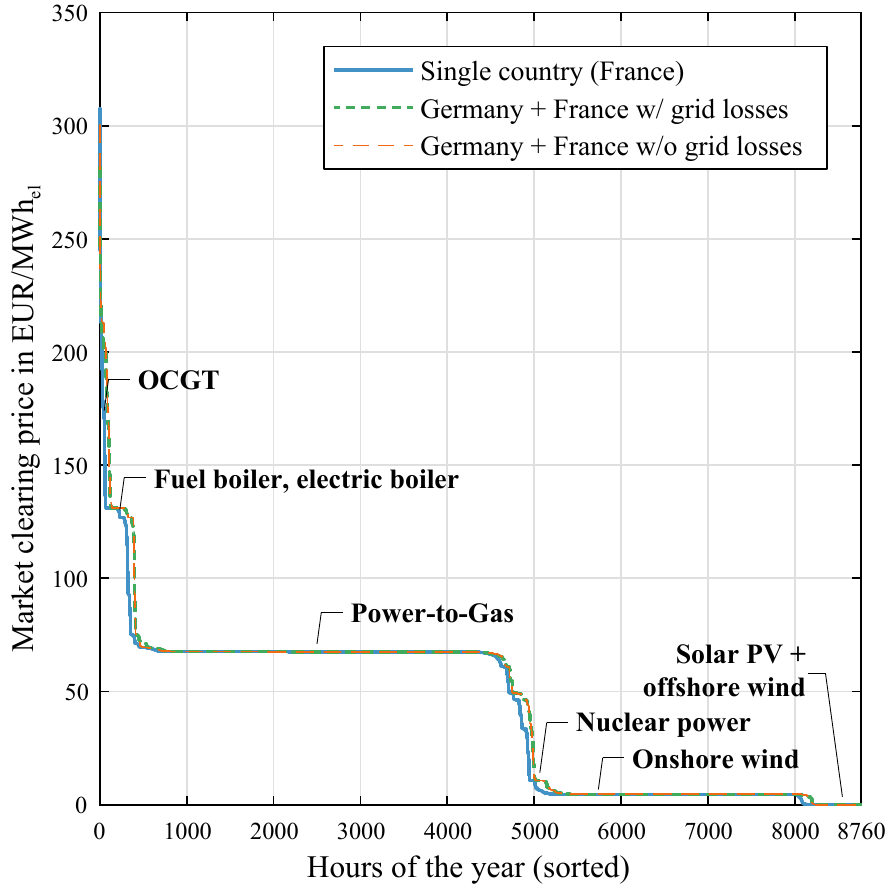}
    \caption{Market clearing price duration curve in the French system with all considered technologies, own illustration based on own computations.}
    \label{Fig.Prices_DEU_FRA}
\end{figure}

The effects of cross-border power exchange between Germany and France on the wholesale electricity market clearing prices are shown for the German and French bidding zones in Fig.~\ref{Fig:price_duration_comparison_deu_fra} and Fig.~\ref{Fig.Prices_DEU_FRA}, respectively.
In a system with only Germany and France, there is a net power export from France to Germany of about 14\,TWh\textsubscript{el}, resulting in a price reduction in Germany and a slight price increase in France.
As we assume an installed capacity of 7.6\,GW\textsubscript{el} of nuclear power plants in France that are still in operation by 2050, there is a price duration curve step at 10.6\,EUR/MWh\textsubscript{el} which is characterised by nuclear power plants setting the price at their marginal costs. Grid losses have only a small effect on the power prices.

\begin{figure}[h!tb]
    \centering\includegraphics[width=\linewidth/2]{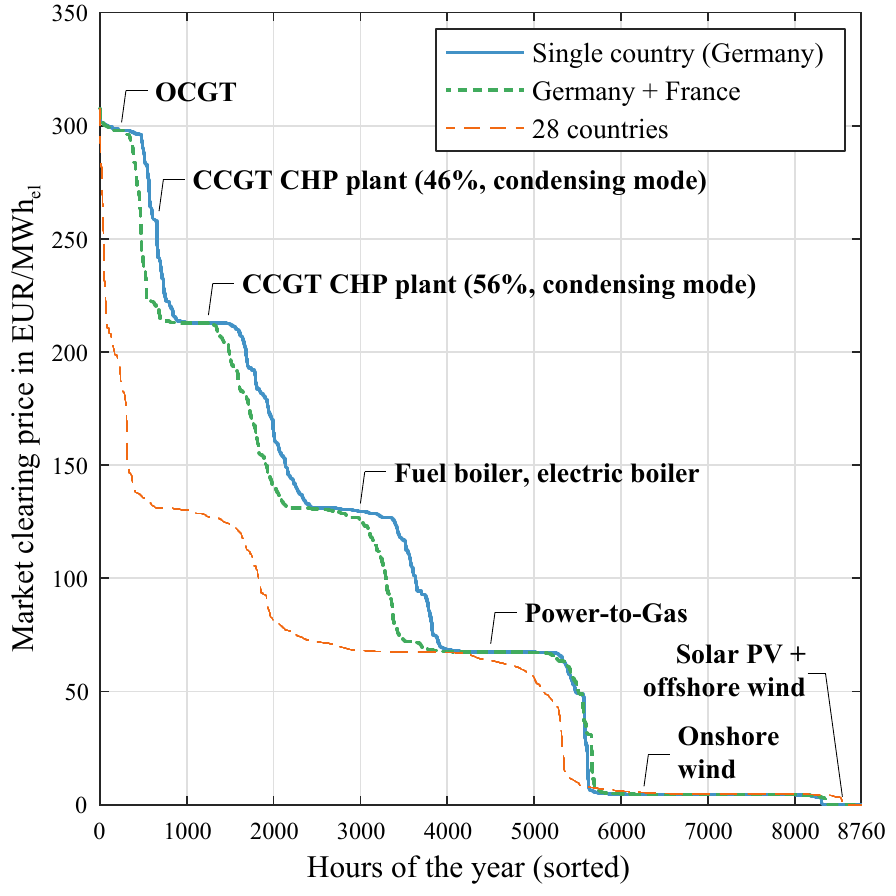}
    \caption{Market clearing price duration curve of Germany with consideration of all 28 European countries in the model, own illustration based on own computations.}
    \label{Fig.Prices_allCountries}
\end{figure}

To highlight the impacts of cross-border integration alongside cross-sector integration in low-carbon electricity markets even more, Fig.~\ref{Fig.Prices_allCountries} shows the effect of cross-border electricity trading if the modelling setup includes all 28 countries, i.e. the EU-27 Member States plus the United Kingdom, Norway, and Switzerland minus Cyprus and Malta.
The possibility to trade electricity across its borders leads to substantially lower prices in the German bidding zone.
Note that Germany is a net importer of electricity with 57\,TWh\textsubscript{el} (this equals 6.3~\% of the power consumption) and that the individual price steps experience even stronger smoothing due to the increased access to flexibility.

\subsection{Market values}

\begin{figure}[h!tb]
    \centering\includegraphics[width=\linewidth/2]{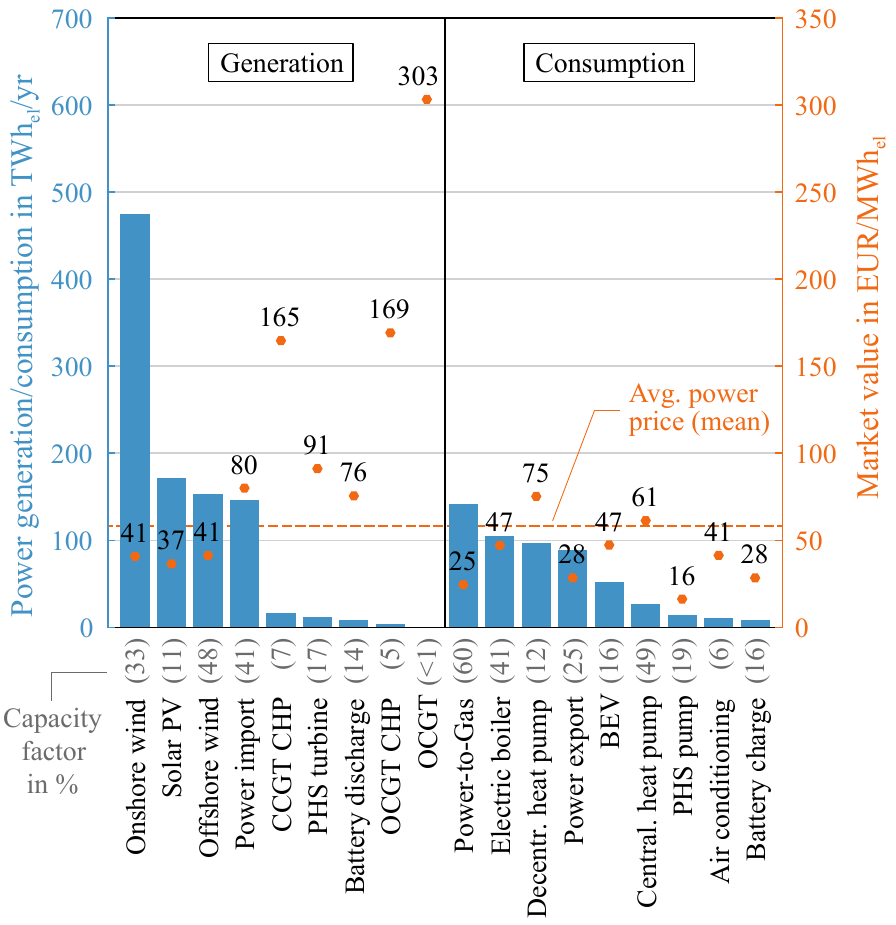}
    \caption{Power generation and consumption and market values (or capture prices) of individual technologies in the wholesale electricity market of Germany in 2050 for the 28 countries scenario, own illustration based on own computations.}
    \label{Fig.Revenues}
\end{figure}

Fig.~\ref{Fig.Revenues} shows the observed market values, also known as capture prices, that the individual generation and consumption technologies capture in the wholesale electricity market under the carbon-neutral scenario for Europe (the 28 countries scenario from above).
Note that the market values are only shown for the German bidding zone.
Moreover, offshore wind generators participate in their home markets and not in a dedicated offshore bidding zone. 
\par
With the average (non-weighted arithmetic mean) electricity price in Germany of about 58\,EUR/MWh\textsubscript{el}, the market value realised by onshore wind of around 41\,EUR/MWh\textsubscript{el} is below that average price.
For solar PV, the captured revenues are even lower with 37\,EUR/MWh\textsubscript{el} due to the high simultaneity of its diurnal availability profile.

\section{Discussion}
\label{S:Discussion}

Based on \citep{Hartel.2020b}, it is essential to point out that the energy system modelling and optimisation framework also exhibits limitations and is based on specific assumptions regarding the representation of markets, policies, and behaviour of actors.
Most notably, the underlying assumptions include perfect competition and absent market power, involving total transparency, rational behaviour of participants, and perfect market efficiency.
In the centralised planner's perspective, strategic bidding behaviour is not considered, and product definitions of electricity market order types are abstracted.
Due to perfect foresight assumptions, future demands are known, and optimal control implies optimistic system operation decisions. 
Moreover, the SCOPE~SD energy system model does not address sequential and strategic decision making structures that emerge in decentralised market environments and are typically captured by market equilibrium models.
Furthermore, the endogenous consideration of uncertainty, if at all, is only achieved by incorporating operational characteristics in large-scale, deterministic, full-year time-series data.
Regarding the observed market clearing prices it is important that we assume two-sided auctions, which allow flexible consumers to make individual bids to the market.
Impacts for other auction or pricing regimes, e.g. pay-as-bid, Vickrey-Clarke-Groves, are not covered here.
\par

In our case study we do not consider demand response of industrial processes which might have higher opportunity costs than the OCGT's generation cost.
For the sake of simplicity, we assume that there is always enough OCGT capacity, implying that industry demand response technology would not be dispatched as it exhibits larger marginal costs than OCGTs.
Even if we restrict the available OCGT capacity to a particular value, industrial demand response's operation times can be expected to be reasonably low.
\par
As mentioned above, we abstract our quantitative analysis from today's regulatory framework so far that electricity or other energy consumption is not associated with additional taxes, fees, levies in the long-term scenario setting.
While these regulatory interventions differ by type and size of market actors, e.g. large industrial or residential end-use consumers, by fuel and end-use sectors, and by jurisdictions in today's market designs, the impacts on opportunity costs and can be substantial.
For instance, see \citep{Hartel.2021} for the impact of a generic cross-sectoral consumer surcharge in low-carbon Europe.
Due to the increasing integration of sectors with new market participants and changing compositions of decentralised and centralised actors, it is expedient to exclude this for this study's purpose.
However, this simplifying decision is not meant to diminish the importance of this complex topic but rather to serve the educational purpose of the stylised and differentiated analysis of market clearing effects of individual technologies. 
\par
The operation of multivalent CHP systems is very flexible due to the hybrid technology setup.
While these system configurations include only already-proven technology components, e.g. a CHP unit, condensing and electric boilers, their combination might require innovative solutions in the technology and regulatory domain, which are assumed to be realised in this context. 
Those CHP systems are only represented by continuous linear formulations of technology clusters, ignoring unit commitment restrictions such as minimum stable generation limits, minimum up- and downtime, or part-load efficiencies.
We further abstract from the provision of ancillary services, e.g. frequency control via balancing markets, which would have an impact on the unit schedules.
Our analysis sees the use of biomass only for process heat and other hard-to-abate end-use demands and applications and not for power generation.
\par
The only available fuel in our scenario is synthetic renewable methane from either domestic production (explicitly modelled) or imports from outside of the considered European system.
We do not consider the use of hydrogen in the power sector but expect the general price formation effects to stay the same in a scenario setting with low-carbon power systems dominated by hydrogen.
Relying on hydrogen ultimately only implies different fuel sourcing costs and conversion efficiencies for the affected technologies, constituting a different point of reference in the resulting price duration levels and steps.
More generally, note that the long-term steady-state of the net-neutral energy system is far from clear.
In other words, whether there will be a world built on hydrogen without emission instruments or a world with the remaining use of fossil fuels and negative emission technologies relying on some form of emission price is an open discussion.
\par
That said, and given the long-term, low-carbon energy system focus of the case study, the resulting drawbacks are a fair compromise to the cross-sectoral interactions that can be taken into account in great detail.

\section{Summary and conclusions}
\label{S:SummaryConclusion}

Due to the modelling framework and the developed long-term scenario, we are able to expose the individual and interdependent price-setting effects in a sector-integrated net-neutral energy system.
We found that the opportunity costs of hybrid cross-sectoral electricity consumers can avoid the penny switching effect of very high and very low prices in power systems primarily based on renewable energy sources.
We still observe some situations with high power prices, occurring when open-cycle or combined-cycle gas turbine, partly as combined heat and power plants, set the price.
When there are only renewable generation bids, i.e. marginal production costs are near-zero, hybrid consumers from the heating sector are mainly responsible for the market clearing price setting. Furthermore, power-to-gas plants exhibit a certain willingness to pay for electricity as long as they can compete with synthetic fuel imports.
All units that can shift energy demand within a limited number of hours, e.g. thermal storage, battery storage, pumped hydro storage, lead to a smoothing of the price step levels.
In the end, there remain only comparably few situations throughout the year in which market-based curtailment of renewable energy sources is unavoidable, leading to situations with very low power prices.
\par
By planning a cost-efficient net-neutral energy scenario, it is shown that market participants with hybrid consumer applications can provide multiple-fuel (and temporal) flexibility to the electricity market, mainly because they integrate variable renewable power generation.
System planners and policy makers should, therefore, push for the implementation of sector integration in reality and enable market operation in a system-friendly way to increase overall welfare in low-carbon power and energy systems.
To that end, harmonising the treatment of fuels in various energy sectors in the context of a suitable market design for low-carbon cross-sectoral energy systems is required.
\par
We confirmed that it is crucial to adequately and explicitly represent cross-sectoral interactions with various technology combinations in energy system models.
This is also highly relevant since the observed market clearing effects in one bidding zone have repercussions in neighbouring bidding zones through the cross-border exchange of electricity, see \citep{Hartel.2021}.
\par
As the key result of our analysis, we see that variable renewable energy sources' market values can be stabilised by the power demand of various direct or indirect electrification applications, including power-to-gas or power-to-heat. Still, their market values lie below the average market prices, potentially resulting in a long-term necessity for support.

\section*{Acknowledgement} 

This paper was financed by Fraunhofer-Gesellschaft zur Förderung der angewandten Forschung e.V. within its programme ``TALENTA speed up''.
We thank Norman Gerhardt for valuable comments and discussions.

\appendix

\section{List of symbols used in SCOPE SD}

\subsection{Model sets and indices}

\begin{compactitem}
\item[$a\in A$]                 Set of stationary (battery) storage system units 
\item[$b\in B$]                 Set of (hybrid) boiler system units 
\item[$g\in G$]                 Set of conventional generation units 
\item[$g\in G^\text{CHP}$]      Set of CHP system units $\left( G^\text{CHP} \subseteq G\right)$
\item[$h\in H$]                 Set of heat pump system units
\item[$i\in I$]                 Set of nodes (bidding zone) 
\item[$j\in J_i$]               Set of nodes connected to node $i$
\item[$l\in L$]                 Set of power-to-fuel (e.g. gas, liquid) units 
\item[$m\in M^\text{COOL}$]     Set of cooling markets
\item[$m\in M^\text{EL}$]       Set of electricity markets 
\item[$m\in M^\text{HEAT}$]     Set of heat markets
\item[$m\in M^\text{ROAD}$]     Set of road transport markets 
\item[$o\in O$]                 Set of cooling system units 
\item[$r\in R$]                 Set of (variable) renewable generation units 
\item[$t\in T$]                 Set of time periods 
\item[$u\in U$]                 Set of (equivalent) hydropower system units 
\item[$v\in V$]                 Set of electric vehicle units 
\item[$\gamma \in \Gamma$]      Set of power producing units, i.e. $\Gamma = G \cup R \cup U \cup A$
\item[$\theta \in\Theta$]       Set of power consuming units, i.e. $\Theta = G^\text{CHP} \cup H \cup U \cup A \cup B \cup V \cup O \cup L$
\item[$(\cdot)_i$]              Subset of $(\cdot)$ belonging to node $i$ (e.g. $G_i, \Gamma_i, \Theta_i$) 
\item[$(\cdot)_m$]              Subset of $(\cdot)$ belonging to market $m$ (e.g. $H_m, \Gamma_m, \Theta_m$) 
\end{compactitem}

\subsection{Decision variables}

All decision variables are continuous and non-negative.\\

\begin{compactitem}
\small{
    \item[$q^\text{CHP}_{g,t}$] CHP heat generation for unit $g$ and time period $t$ (MW\textsubscript{th})
    \item[$q^\text{IN}_{(\cdot),t}$] Thermal storage input for unit $g$, $o$ and time period $t$ (MW\textsubscript{th})
    \item[$q^\text{OUT}_{(\cdot),t}$] Thermal storage output for unit $g$, $o$ and time period $t$ (MW\textsubscript{th})
    \item[$q^\text{S}_{(\cdot),t}$] Thermal storage level for unit $g$, $o$ and time period $t$ (MWh\textsubscript{th})
    \item[$x_{i,j,t}$] Power transfer from node i to node j in time period $t$ (MW\textsubscript{el})
    \item[$x^\text{CON}_{(\cdot),t}$] Power consumption for unit $(\cdot)$ (e.g. $h$) and time period $t$ (MW\textsubscript{el})
    \item[$x^\text{CON,HP}_{h,t}$] Power consumption of the heat pump for unit $h$ and time period $t$ (MW\textsubscript{el})
    \item[$x^\text{CON,BU}_{h,t}$] Power consumption of the electric backup unit for unit $h$ and time period $t$ (MW\textsubscript{el})
    \item[$x^\text{CU}_{r,t}$]  Power curtailment for renewable generation unit $r$ and time period $t$ (MW\textsubscript{el})
    \item[$x^\text{GEN}_{(\cdot),t}$] Power generation for unit $(\cdot)$ (e.g. $g$) and time period $t$ (MW\textsubscript{el})
    \item[$x^\text{GEN,PT}_{u,t}$]  Pumped-turbine generation for hydropower unit $u$ and time period $t$ (MW\textsubscript{el})
    \item[$x^\text{GEN,T}_{u,t}$]  Turbine generation for hydropower unit $u$ and time period $t$ (MW\textsubscript{el})
    \item[$x^\text{IN}_{(\cdot),t}$]  Electricity storage input for unit $a,v$ and time period $t$ (MW\textsubscript{el})
    \item[$x^{\text{LC}+/-}_{(\cdot),t}$] Positive/negative load change for unit $(\cdot)$ (e.g. $g$) and time period $t$ (MW\textsubscript{el})
    \item[$x^\text{OUT}_{(\cdot),t}$]  Electricity storage output for unit $a,v$ and time period $t$ (MW\textsubscript{el})
    \item[$x^\text{S}_{(\cdot),t}$]  Electricity storage level unit $a,v$ and time period $t$ (MWh\textsubscript{el})
    \item[$x^\text{S}_{u,t}$]  Conventional reservoir storage level for hydropower unit $u$ and time period $t$ (MWh\textsubscript{el})
    \item[$x^\text{S,P}_{u,t}$]  Pumped reservoir storage level for hydropower unit $u$ and time period $t$ (MWh\textsubscript{el})
    \item[$x^\text{SP}_{u,t}$]  Conventional reservoir spillage for hydropower unit $u$ and time period $t$ (MW\textsubscript{el})
    \item[$x^\text{SP,P}_{u,t}$]  Pumped reservoir spillage for hydropower unit $u$ and time period $t$ (MW\textsubscript{el})
    \item[$y^\text{CB}_{(\cdot),t}$] Condensing boiler fuel consumption for unit $g,b$ and time period $t$ (MW\textsubscript{th})
    \item[$y^\text{CON}_{(\cdot),t}$] (Total) fuel consumption for unit $(\cdot)$ (e.g. $g$) and time period $t$ (MW\textsubscript{th})
    \item[$y^{\text{GEN}}_{l,t}$] Fuel generation for unit $l$ and time period $t$ (MW\textsubscript{th})
    \item[$z_{v,t}^{\text{EL}}$] Distance driven by electric drive of vehicle unit $v$ (km/h)
    \item[$z_{v,t}^{\text{ICE}}$] Distance driven by internal combustion engine of vehicle unit $v$ (km/h)
}
\end{compactitem}

\subsection{Parameters}

\begin{compactitem}
\small{
    \item[$AV_{(\cdot),t}$]         Availability for unit $(\cdot)$ and time period $t$ (\% of installed capacity)
    \item[$C_{(\cdot)}^\text{CB}$]  Variable boiler production cost for hybrid boiler unit $b$ or CHP unit $g$ (EUR/MWh\textsubscript{th})  
    \item[$C_{g}^\text{CHP}$]  Variable production cost for heat extraction for CHP unit $g$ (EUR/MWh\textsubscript{el})
    \item[$C_{r}^\text{CU}$]        Curtailment cost for unit $r$ (EUR/MWh\textsubscript{el})
    \item[$C_{v}^\text{ICE}$]  Variable internal combustion engine cost for vehicle unit $v$ (EUR/MWh\textsubscript{th})
    \item[$C_{(\cdot)}^\text{LC}$]  Load change cost for unit $(\cdot)$ (EUR/MW\textsubscript{el})
    \item[$C_{(\cdot)}^\text{VP}$]  Variable production cost for unit $(\cdot)$ (EUR/MWh\textsubscript{el})
    \item[$D_{m,t}$]                Demand for market type $m$ and time period $t$ (e.g. MWh\textsubscript{el}/h, MWh\textsubscript{th}/h)
    \item[$EC_{v}$] Distance-specific electricity consumption of vehicle unit $v$ (MWh\textsubscript{el}/km)
    \item[$FC_{v}$] Distance-specific fuel consumption of vehicle unit $v$ (MWh\textsubscript{th}/km) 
    \item[$FS_{v}$] Flexible charging share for vehicle unit $v$ (1
    \item[$NTC_{i,j,t}$]            Net transfer capacity for power exchange between nodes $i$ and $j$ and time period $t$ (MWh\textsubscript{el}/h)
    \item[$P^{CH_4}$]         Price for (renewable) methane (EUR/MW\textsubscript{fuel})
    \item[$PF_{l}$]         Power-to-fuel conversion factor for unit $l$ (MWh\textsubscript{th}/MWh\textsubscript{el})
    \item[$PH_{g}$]         Power-to-heat ratio (backpressure limit) factor for unit $g \in G^\text{CHP}$ (MWh\textsubscript{el}/MWh\textsubscript{th})
    \item[$PL_{g}$]         Power loss factor for unit $g \in G^\text{CHP}$ (MWh\textsubscript{el}/MWh\textsubscript{th})
    \item[$ST_{(\cdot),t}$]         Solar thermal contribution factor of unit $(\cdot)$ (1)
    \item[$TL_{i,j}$]                Linear power transmission loss factor for power exchange between nodes $i$ and $j$ (1)
    \item[$\underbarnew{X}^\text{CON}_{(\cdot),t}$]  Min. power consumption limit for unit $(\cdot)$ (e.g. $h$) and time period $t$ (MW\textsubscript{el})
    \item[$\overbar{X}^\text{CON,FC}_{v,t}$] Max. flexible charging profile of vehicle unit $v$ and time period $t$ (MWh\textsubscript{el}/h)
    \item[$X_{v,t}^\text{CON,IC}$] Inflexible (fixed) charging profile of vehicle unit $v$ and time period $t$ (MWh\textsubscript{el}/h)
    \item[$\underbarnew{X}^\text{GEN}_{(\cdot),t}$]  Min. power generation limit for unit $(\cdot)$ (e.g. $g$) and time period $t$ (MW\textsubscript{el})
    \item[$\overbar{X}^\text{GEN}_{(\cdot)}$]  Power generation capacity for unit $(\cdot)$ (e.g. $g$) (MW\textsubscript{el})
    \item[$\overbar{X}^\text{GEN,PT}_{u,t}$]  Max. pumped-turbine generation for hydropower unit $u$ and time period $t$ (MW\textsubscript{el})
    \item[$\overbar{X}^\text{GEN,T}_{u,t}$]  Max. turbine generation for hydropower unit $u$ and time period $t$ (MW\textsubscript{el})
    \item[$X^\text{N}_{u,t}$] Equivalent natural energy inflow into main reservoir for hydropower unit $u$ and time period $t$  (MW\textsubscript{el}) 
    \item[$X^\text{N,P}_{u,t}$] Equivalent natural energy inflow into pumped reservoir for hydropower unit $u$ and time period $t$ (MW\textsubscript{el})
    \item[$\underbarnew{X}^{\text{S}}_{v,t}$] Min. state-of-charge profile of vehicle unit $v$ and time period $t$ (MWh\textsubscript{el})
    \item[$\overbar{X}^{\text{S}}_{v,t}$] Max. state-of-charge profile of vehicle unit $v$ and time period $t$ (MWh\textsubscript{el})
    \item[$\overline{Z}^{\text{EL}}_{v,t}$] Max. distance driven by electric drive of vehicle unit $v$ and time period $t$ (km/h)
    \item[$\eta^\text{CB}_{(\cdot)}$] Condensing boiler efficiency for unit $g,b$ (1)
    \item[$\eta^\text{CH}_{v}$] Charging efficiency for vehicle unit $v$ (1)
    \item[$\eta^\text{CON}_{(\cdot)}$] Electric heating/cooling efficiency for unit $g$, $b$, $o$ (MWh\textsubscript{th}/MWh\textsubscript{el})
    \item[$\eta^\text{CON}_{h,t}$] Electric heating efficiency for unit $h$ and time period $t$ (MWh\textsubscript{th}/MWh\textsubscript{el})    
    \item[$\eta^\text{GEN}_{g}$] Efficiency of generation unit $g$
    \item[$\eta^\text{IN}_{(\cdot)}$] Storage input efficiency for unit $(\cdot)$ (e.g. $g$, $a$) (1)
    \item[$\eta^\text{OUT}_{(\cdot)}$] Storage output efficiency for unit $(\cdot)$ (e.g. $g$, $a$) (1)
    \item[$\eta^\text{P}_u$] Pump efficiency factor for hydropower unit $u$ (1)
    \item[$\lambda_{(\cdot)}^\text{S}$] Linear thermal or electricity storage loss factor for unit $(\cdot)$ (e.g. $g$, $a$) (1)
    \item[$\Lambda_{(\cdot)}^\text{S}$] Linear thermal or electricity storage self-discharge factor for unit $(\cdot)$ (e.g. $g$, $a$) (1)    
    \item[$\Pi^\text{BU}_{(\cdot)}$] Design factor of maximum thermal backup boiler output for units $g$ and $b$ (1)
    \item[$\Pi^\text{CHP}_{g}$] Design ratio factor of maximum thermal chp output in relation to the maximum heat demand for CHP unit $g$ (1)
    \item[$\Pi^\text{CON}_{g}$] Design factor of maximum electric backup boiler output for CHP unit $g$ (1) 
    \item[$\Phi_{v}$] Market share of vehicle unit $v$ (1)
    \item[$\chi_{(\cdot)}$] Linear heating network transmission loss factor for unit $(\cdot)$ (1) 
}
\end{compactitem}

\section{Additional tables for CHP opportunity cost calculation}
\label{Appendix.tables}

Table \ref{Tab.OppcostBoiler} explains the derivation of the opportunity costs between CHP unit and fuel boiler. A marginal increase of the power supply of the CHP unit of $\epsilon$ can lead to an increased heat supply of $\frac{\epsilon}{PH_{g}}$ according to the power-to-heat ratio $PH_{g}$. This results in additional costs for the power generation ($\epsilon \cdot \frac{P^{CH_4}}{\eta^\text{GEN}_g}$) and the heat supply ($\frac{\epsilon}{PH_{g}} \cdot P^{CH_4} \cdot PL_{g} $) accouring to \eqref{eq:chpCosts}. On the other hand the fuel boiler could decrease its heat supply by the same extent of the increase of the heat supply of the CHP system resulting in a cost reduction of $\frac{\epsilon}{PH_{g}} \cdot \frac{P^{CH_4}}{\eta^\text{CB}_{g}} $. Therefore, the opportunity costs of a switch of heat supply between CHP unit and fuel boiler are $ \frac{P^{CH_4}}{\eta^\text{GEN}_g} + \frac{1}{PH_{g}} \cdot P^{CH_4} \cdot PL_{g}-\frac{1}{PH_{g}} \cdot \frac{P^{CH_4}}{\eta^\text{CB}_{g}} $ as given in \eqref{eq:cost_Boiler_marginal}.

\begin{table*}[h!tb]
\footnotesize
\centering
\begin{tabular}{l | c c |c}
\hline
\textbf{} & \textbf{Power generation in $\text{MWh}_{\text{el}}$} & \textbf{Heat in $\text{MWh}_\text{th}$} & \textbf{Costs in EUR}\\
\hline
CHP & $+\epsilon $ & $+\frac{\epsilon}{PH_{g}}$ & $\epsilon \cdot \frac{P^{CH_4}}{\eta^\text{GEN}_g} + \frac{\epsilon}{PH_{g}} \cdot P^{CH_4} \cdot PL_{g} $ \\
Fuel boiler & 0 & $-\frac{\epsilon}{PH_{g}}$ & $-\frac{\epsilon}{PH_{g}} \cdot \frac{P^{CH_4}}{\eta^\text{CB}_{g}} $ \\
\hline
Sum &  $+\epsilon$ & 0 & $\epsilon \cdot \frac{P^{CH_4}}{\eta^\text{GEN}_g} + \frac{\epsilon}{PH_{g}} \cdot P^{CH_4} \cdot PL_{g}-\frac{\epsilon}{PH_{g}} \cdot \frac{P^{CH_4}}{\eta^\text{CB}_{g}} $\\
\hline
\end{tabular}
\caption{Calculation of opportunity costs of fuel boiler heat supply.}
\label{Tab.OppcostBoiler}
\end{table*}

The calculation of opportunity costs between CHP unit and electric backup unit is even more complex and explained in Table \ref{Tab.OppcostP2H}. The change for the CHP unit is identical to the aforementioned opportunity between the CHP unit and the fuel boiler. The balancing of the heat supply with the electric backup ($-\frac{\epsilon}{PH_{g}} $) leads to less power consumption according to the efficiency $\eta^\text{CON}_g$. Therefore, indirectly power generation can be increased by $\frac{\epsilon}{PH_{g} \cdot \eta^\text{CON}_g}$. There are no costs associated to the electric backup (abstracting from today's regulatory framework). In total the opportunity costs of a switch of heat supply between CHP unit and electric backup unit are 
$\frac{\frac{P^{CH_4}}{\eta^\text{GEN}_g}}{1+\frac{1}{PH_{g}\cdot \eta^\text{CON}_g}} + \frac{P^{CH_4}\cdot \frac{PL_{g}}{PH_{g}}}{1+\frac{1}{PH_{g}\cdot\eta^\text{CON}_g}} $ as given in \eqref{eq:cost_P2H_marginal}.

\begin{table*}[h!tb]
\footnotesize
\centering
\begin{tabular}{l | c c |c}
\hline
\textbf{} & \textbf{Power generation in $\text{MWh}_{\text{el}}$} & \textbf{Heat in $\text{MWh}_\text{th}$} & \textbf{Costs in EUR}\\
\hline
CHP & $+\epsilon $ & $+\frac{\epsilon}{PH_{g}}$ &  $\epsilon \cdot \frac{P^{CH_4}}{\eta^\text{GEN}_g} + \frac{\epsilon}{PH_{g}} \cdot P^{CH_4} \cdot PL_{g} $ \\
Electric boiler & $-(-\frac{\epsilon}{PH_{g} \cdot \eta^\text{CON}_g} )$ & $-\frac{\epsilon}{PH_{g}} $ & 0 \\
\hline
Sum &  $\epsilon +\frac{\epsilon}{PH_{g} \cdot \eta^\text{CON}_g} $ & 0 & $\epsilon \cdot \frac{P^{CH_4}}{\eta^\text{GEN}_g} + \epsilon \cdot P^{CH_4} \cdot \frac{PL_{g}}{PH_{g}} $\\
\hline
\end{tabular}
\caption{Calculation of opportunity costs of electric heat supply.}
\label{Tab.OppcostP2H}
\end{table*}

Table \ref{TabCHPcases} gives an overview of all cases for the price setting of multivalent CHP systems where $\eta^\text{GEN}_{OCGT}$ is the electrical efficiency of the OCGT.
For case G the derivation of the heat value can be found in Table \ref{Tab.CHP-CaseG}.

\begin{table*}[h!tb]
\footnotesize
\centering
\resizebox{\textwidth}{!}{%
\begin{tabular}{p{0.35cm} p{1.6cm} p{1.8cm} p{1.3cm} p{0.8cm} p{0.8cm} p{2cm} p{2cm} p{1.4cm} }
\textbf{Case} & \textbf{Power system state} & \textbf{CHP system state} & \textbf{CHP system} & \textbf{Fuel boiler} & \textbf{Electric backup}  & \textbf{Heat value setting} & \textbf{Power price setting} & \textbf{Frequency in h/yr} \\
\hline \hline
A & Power shortage, OCGT running & CHP on backpressure-line with maximum boiler output & full load & partial load & full load & $\frac{P^{CH_4}}{\eta^\text{CB}_{g}}=128.2$ (gas boiler costs) & OCGT opportunity & \begin{flushright} 123 \end{flushright} \\ \hline
B & Power shortage, OCGT not running & CHP operates on backpressure-line & partial load & partial load & full load &  $\frac{P^{CH_4}}{\eta^\text{CB}_{g}}=128.2$ (gas boiler costs)  & Opportunity between CHP and fuel boiler, see \eqref{eq:cost_Boiler_marginal} &\begin{flushright}  10\end{flushright} \\ \hline
C & Power surplus, OCGT not running & CHP offline; el. backup in full load  & not running & partial load & full load & $\frac{P^{CH_4}}{\eta^\text{CB}_{g}}=128.2$ (gas boiler costs) & RES curtailment & \begin{flushright} 1,232 \end{flushright} \\ \hline
D & Power shortage, OCGT running & CHP on backpressure-line with maximum boiler output & full load & not running & partial load & power price/$\eta^\text{CON}_g$ (el. backup opportunity) & OCGT opportunity & \begin{flushright} 377\end{flushright} \\ \hline
E & Power shortage, OCGT not running & CHP operates on backpressure-line & partial load & not running & partial load & power price/$\eta^\text{CON}_g$ (el. backup opportunity) & Opportunity between CHP and el. backup, see \eqref{eq:cost_P2H_marginal} & \begin{flushright} 49\end{flushright} \\ \hline
F & Power surplus, OCGT not running & CHP offline; el. backup in partial load & not running & not running & partial load & power price/$\eta^\text{CON}_g$ (el. backup opportunity) & RES curtailment &\begin{flushright}  5,565 \end{flushright} \\ \hline
G & Power shortage, OCGT running & CHP with maximum boiler output & full load & not running & not running & $P^{CH_4}\cdot PL_{g} \cdot (1+1/\eta^\text{GEN}_{OCGT}) - 1/\eta^\text{GEN}_g) $ (s. Table \ref{Tab.CHP-CaseG}) & OCGT opportunity & \begin{flushright} 1,341 \end{flushright}\\ \hline 
H & Power shortage, OCGT not running & CHP operates below backpressure-line & partial load & not running & not running & $P^{CH_4}\cdot PL_{g}$ (heat supply costs of CHP system) & CHP opportunity (marginal costs) & \begin{flushright} 63\end{flushright} \\ \hline
\end{tabular}
}
\caption{Overview of different cases for opportunity costs of multivalent CHP systems.}
\label{TabCHPcases}
\end{table*}

\begin{table*}[h!tb]
\footnotesize
\centering
\begin{tabular}{l | c c |c}
\hline
\textbf{} & \textbf{Power generation in $\text{MWh}_{\text{el}}$} & \textbf{Heat in $\text{MWh}_\text{th}$} & \textbf{Costs in EUR}\\
\hline
CHP & $-\epsilon \cdot PL_{g} $ & $+\epsilon$ &  $\epsilon \cdot P^{CH_4}\cdot PL_{g} -\epsilon \cdot P^{CH_4} \cdot \frac{PL_{g}}{\eta^\text{GEN}_g}$ \\
OCGT & $+\epsilon \cdot PL_{g}$ & 0 & $\epsilon \cdot P^{CH_4}\cdot \frac{PL_{g}}{\eta^\text{GEN}_{OCGT}}$ \\
\hline
Sum &  0 & $+\epsilon$ & $ \epsilon \cdot P^{CH_4}\cdot PL_{g} -\epsilon \cdot P^{CH_4} \cdot \frac{PL_{g}}{\eta^\text{GEN}_g}+\epsilon \cdot P^{CH_4}\cdot \frac{PL_{g}}{\eta^\text{GEN}_{OCGT}}$\\
\hline
\end{tabular}
\caption{Calculation of opportunity costs of electric heat supply.}
\label{Tab.CHP-CaseG}
\end{table*}

\bibliographystyle{model2-names}
\bibliography{manuscript_Boettger.bib}

\end{document}